\documentclass[journal]{IEEEtranTCOM}

\normalsize

\usepackage[dvips]{graphicx} 
\usepackage{amsmath,amssymb,amsfonts}
\usepackage{algorithm}
\usepackage{algpseudocode}
\usepackage{graphicx}
\usepackage{textcomp}
\usepackage{xcolor}
\usepackage{amsmath}
\usepackage{relsize}
\usepackage{amssymb}
\usepackage[utf8]{inputenc}
\usepackage[sorting=none,style=ieee]{biblatex}
\usepackage{nccmath}
\usepackage{ragged2e}
\usepackage[caption=false,font=normalsize,labelfont=sf,textfont=sf]{subfig}
\usepackage{balance}
\usepackage{float}
\usepackage{booktabs} 
\usepackage{mathtools}

\usepackage{multirow}

\let\labelindent\relax
\usepackage{enumitem}
\newlist{steps}{enumerate}{1}
\setlist[steps, 1]{label = Step \arabic*:}

\graphicspath{{Figures/}}

\newcommand{\CC}{\mathcal{C}}
\newcommand{\TC}{\mathcal{T}}
\newcommand{\LC}{\mathcal{L}}
\DeclareMathOperator*{\argmin}{arg\,min}
\DeclareUnicodeCharacter{2212}{-}
\DeclareUnicodeCharacter{0301}{\'{e}}

\algrenewcommand\algorithmicrequire{\textbf{Inputs:}}
\algrenewcommand\algorithmicensure{\textbf{Output:}}

\def\BibTeX{{\rm B\kern-.05em{\sc i\kern-.025em b}\kern-.08em
    T\kern-.1667em\lower.7ex\hbox{E}\kern-.125emX}}
    
\AtBeginBibliography{\small}

\begin{document}

\title{Ordered Reliability Direct Error Pattern Testing Decoding Algorithm}
\author{Reza Hadavian, Xiaoting Huang, Dmitri Truhachev, Kamal El-Sankary, Hamid Ebrahimzad and Hossein Najafi
\thanks{This work has been submitted to the IEEE for possible publication. Copyright may be transferred without notice, after which this version may no longer be accessible.}
\thanks{This work was supported in part by NSERC Discovery Grant. The material of this paper is presented in part at the IEEE Globecom, Kuala-Lumpur, Malaysia, Dec. 2023.}
\thanks{Reza Hadavian, Xiaoting Huang, Dmitri Truhachev and Kamal El-Sankary are with Department of Electrical and Computer Engineering, Dalhousie University, Halifax, Canada (e-mail: reza.hadavian@dal.ca; xiaoting.huang@dal.ca; dmitry@dal.ca; kamal.el-sankary@dal.ca). }
\thanks{Hamid Ebrahimzad and Hossein Najafi are with Huawei Corporation, Ottawa, Canada (e-mail: hamid.ebrahimzad@huawei.com; hossein.najafi@huawei.com).}
}


\maketitle

\begin{abstract}
We introduce a novel universal soft-decision decoding algorithm for binary block codes called ordered reliability direct error pattern testing (ORDEPT). 
Our results, obtained for a variety of popular short high-rate codes, demonstrate that ORDEPT outperforms state-of-the-art decoding algorithms of comparable complexity such as ordered reliability bits guessing random additive noise decoding (ORBGRAND) in terms of the decoding error probability and latency.
The improvements carry on to the iterative decoding of 
product codes and convolutional product-like codes, where we present a new adaptive decoding algorithm and demonstrate the ability of ORDEPT to efficiently find multiple candidate codewords to produce soft output.
\end{abstract}

\begin{IEEEkeywords}
Forward error correction, GRAND, ORBGRAND, Chase, DEPT
\end{IEEEkeywords}

\IEEEpeerreviewmaketitle

\section{Introduction}

Short codes are crucial for modern communications. They are used to encode short packets in wireless systems for ultra-reliable low-latency communications (URLLC), machine-to-machine (M2M) communications, self-driving vehicles, internet-of-things (IoT), etc. \cite{durisi2016toward,she2017radio,chen2018ultra}. They also serve as component codes in more complex forward error-correction (FEC) code structures such as generalized low-density parity-check (LDPC) codes \cite{tanner1981recursive}, staircase~\cite{smith2011staircase}, zipper~\cite{sukmadji2019zipper}, and open FEC (oFEC)~\cite{sluysk2019open} codes. Short codes are also utilized as inner codes in concatenated optical FECs such as the 400ZR fiber-optical communication standard~\cite{400ZR}. 



Since the early days of coding theory, there has been a significant focus on the development of accurate decoding algorithms. Initially, decoders with an algebraic, hard-input hard-output approach were devised, capitalizing on the algebraic structure of the code~\cite{berlekamp2015algebraic}. In 1972, Chase introduced a decoding algorithm that remained indifferent to the specific code structure and instead made use of soft channel output information~\cite{chase1972class}. This algorithm employs a set of patterns designed to introduce perturbations to bit positions with low reliability, specifically those characterized by the lowest absolute log-likelihood ratios (LLRs). Subsequently, it performs hard-decision decoding to identify potential candidate codewords. The decoder outputs a candidate codeword that demonstrates the highest reliability, quantified in terms of Euclidean distance to the received sequence or the sum of absolute LLRs across distinct bit positions. Over time, numerous research contributions have emerged, aiming to generalize and enhance the performance of the Chase decoder \cite{tendolkar1984generalization, tokushige2001selection, tokushige2010test}. It is important to note that the complexity of the algorithm increases with the number of test patterns used, as each pattern necessitates a hard-decision decoding attempt.

A method known as direct error-pattern testing (DEPT) was recently introduced in~\cite{truhachev2020efficient}. DEPT chooses a set of partial error patterns (PEPs) using the positions of the lowest absolute LLRs and then finalizes each PEP by identifying the last error position using the syndrome computed from the hard-decision received sequence. The approach relies on a list of PEPs that play the most significant role in contributing to the error probability.

The ordered statistics decoding (OSD) algorithm~\cite{fossorier1995soft}, offers a soft decoding approach alternative to Chase decoding. OSD initiates the decoding from the most reliable bit positions in contrast to the Chase algorithm, which commences from the least reliable ones. OSD essentially re-encodes sequences generated through test patterns and assesses them in terms of their reliability. Nevertheless, the computational complexity of OSD increases exponentially with the dimension of the code, making it better suited for lower-rate codes.

A novel strategy known as Guessing Random Additive Noise Decoding (GRAND), which was recently introduced in \cite{duffy2019capacity}, offers a practical solution to achieving near Maximum Likelihood (ML) performance for short high-rate binary codes operating on a binary symmetric channel (BSC). Diverging from the conventional decoding methods that rely on exploiting the structural characteristics of the code, GRAND places its focus on pinpointing the most probable noise effect. It accomplishes this by utilizing error patterns that are sorted in a descending order of likelihood and assessing whether the modulo-2 addition of an error pattern and the hard-decision received sequence corresponds to a valid codeword in the codebook. The first codeword member discovered through this process is chosen as the decoded codeword. The GRAND decoding approach has gained a considerable attention from researchers due to several advantageous features. Firstly, GRAND is a universal decoder in a sense that is does not exploit the structure of the code apart from its parity-check matrix used for syndrome computation, and can be employed for any linear code. Furthermore, it lends itself well to parallelization, enabling the simultaneous testing of multiple error patterns. To further reduce latency, an enhanced version of GRAND, referred to as GRAND with abandonment (GRANDAB), was introduced in~\cite{duffy2019capacity}, which terminates the codeword search after a predefined number of failed queries.

The universality of GRAND has prompted a re-evaluation of conventional error-correction codes, such as cyclic redundancy check (CRC) codes, which were traditionally utilized for error detection applications. Moreover, GRAND is capable of efficiently decoding CRC-assisted polar (CA-Polar) codes \cite{channel2009Arikan}, a coding scheme that is used in control channel communications within the 5G New Radio standard \cite{3gpp2020technical}.

In addition to the original hard-decision GRAND, a soft maximum likelihood decoder named soft GRAND (SGRAND) was introduced~\cite{solomon2020soft}. However, the implementation of SGRAND in hardware presents formidable challenges due to its inherent NP complexity. To address this complexity problem, the ordered reliability bits GRAND (ORBGRAND) that employs a linear approximation of the ordered absolute LLR distributions used to order the error patterns was introduced in~\cite{duffy2022ordered}. While accurate piece-wise linear approximations of these distributions within multi-line ORBGRAND enhance decoding results, they also introduce additional complexity by utilizing multiple signal-to-noise ratio (SNR)-dependent lists of queries. Furthermore, to streamline the process of pattern generation in ORBGRAND, two alternative approaches, namely QGRAND and DSGRAND, were proposed in \cite{yuan2022role}. Another strategy, LGRAND, aims to bridge the gap between ORBGRAND and SGRAND by creating a list of multiple candidate codewords and ultimately selecting the one with the highest likelihood. All the above-mentioned ORBGRAND variants  necessitate testing of a long list of error patterns (queries) until a valid codeword is identified. This process may result in long average and worst-case latency, rendering these algorithms potentially unsuitable for high-speed applications such as optical FECs.

In~\cite{10089737}, the feasibility of employing ORBGRAND as a component decoder in product codes, typically decoded using the Chase-Pyndiah algorithm~\cite{pyndiah1998near}, is investigated. This exploration focuses on a product code with Bose-Chaudhuri-Hocquenghem (BCH)~\cite{hocquenghem1959codes, bose1960class} $\mathrm{BCH}(32,26,4)$ component codes. The study illustrates that ORBGRAND outperforms its Chase-Pyndiah counterpart when both employ the same number of codewords in their lists. However, for longer codes such as $\mathrm{BCH}(256,239)$, which are utilized in the oFEC standard~\cite{sluysk2019open} used for high-speed fiber-optical communications, ORBGRAND encounters challenges in identifying even the second candidate codeword. Since a competitor codeword is essential for generating meaningful soft output \cite{condo2022iterative}, its absence results in a significant degradation in the bit error rate (BER), since the reliability of the decoded codeword bits needs to be estimated empirically~\cite{condo2022iterative}. While a recent implementation proposal of ORBGRAND called step-GRAND~\cite{abbas2023step} aims to reduce its latency, it does not contribute to enhancing the performance of iterative decoding  either. To address the challenge of identifying multiple candidate codewords, \cite{condo2022iterative} suggests seeking a competitor codeword through Chase decoding. However, this solution compromises the universal applicability of the ORBGRAND decoder and adds to its complexity. Another proposal outlined in \cite{condo2022iterative} involves employing a larger number of test patterns, which also leads to increased worst-case latency.

In this paper, we introduce a novel decoding algorithm that is able to outperform the existing approaches of similar complexity in terms of the decoding error probability and latency. Our contributions are as follows:
\begin{itemize}
\item We propose an algorithm called ordered reliability direct error pattern testing (ORDEPT), which completes a set of pre-defined partial error patterns (PEP)s ordered according to their reliability. The algorithm is able to efficiently complete the patterns and test them in parallel batches (shots). 
\item We present a number of improvements to ORDEPT such as finding and utilizing multiple successful queries, thresholding to reduce latency, simplified completion function for BCH codes to reduce complexity and save power in a hardware implementation. We also develop a generalization (ORDEPT$x$, $x>1$), where PEPs are completed by upto $x$ error positions instead of just one. 
\item We propose a new adaptive iterative soft-decision decoding algorithm based on ORDEPT candidate codeword metrics and show that it outperforms the original adaptive approach~\cite{deng2023adaptive}.
\end{itemize}
We demonstrate that ORDEPT can achieve a superior error-correction performance in comparison to ORBGRAND, with a significantly smaller list of queries that leads to reduced latency. 
Our proposed decoder maintains the universality and parallelizability properties.
Moreover, ORDEPT efficiently finds multiple candidate codewords while using a reduced number of queries which is essential for soft input soft output (SISO) decoding needed in turbo decoding systems. This has been demonstrated by employing it as a component decoder in product codes and convolutional product codes, specifically the oFEC. For the case of product-like codes, the advantages of ORDEPT over the traditional Chase~II algorithms of comparable complexity can also be pointed out.

The rest of this paper is organized as follows. The system model is introduced in Section~\ref{sec:SystemModel} along with the proposed ORDEPT algorithm. Section~\ref{sec:Component} describes its application as a component decoder passing soft information. Section~\ref{sec:SimulationResults} presents numerical results for various individual and product-like codes. Finally, Section~\ref{sec:conclusion} concludes the paper. 


\section{System Model and Decoding Algorithms} \label{sec:SystemModel}

\subsection{System and Channel Model} \label{subsec:ChannelModel}
Consider a binary information vector of length $k$, which undergoes encoding with a binary linear block code denoted by $\CC$. This encoding process yields an $n$-bit codeword, represented by $\boldsymbol{c}=({c}_1,{c}_2,\ldots,{c}_n)$, where each ${c}_i$ takes on values from the set $\{0,1\}$ for $i=1,\ldots,n$. The codeword $\boldsymbol{c}$ is then modulated by binary phase-shift keying (BPSK) modulator. Each codeword component $c_i$ is mapped to $x_i\in \{-1,1\}$ via $x_i=2c_i-1$. Subsequently, the modulated signal $\boldsymbol{x}=({x}_1,{x}_2,\ldots,{x}_n)$ is transmitted through a real-valued additive white Gaussian noise (AWGN) channel. The resulting channel output can be expressed via
\begin{equation} \label{eq:channel_output}
    \boldsymbol{y}=\boldsymbol{x}+\boldsymbol{\zeta},
\end{equation}
where $\boldsymbol{\zeta}$ represents a vector composed of independent and identically distributed (i.i.d.) Gaussian random variables characterized by a zero mean and a variance of $\sigma^2$. Utilizing Equation~\ref{eq:channel_output}, the hard-decision received sequence $\boldsymbol{w}$ can be computed as $\boldsymbol{w}=\frac{1}{2}(\mathrm{sign}(\boldsymbol{y})+1)$. The noise effect, denoted by  ${\boldsymbol{z}}$, is then defined as the disparity between the transmitted codeword $\boldsymbol{c}$ and the hard-decision sequence $\boldsymbol{w}$. This disparity is expressed as ${\boldsymbol{z}}=\boldsymbol{c}\oplus\boldsymbol{w}$.

Let us denote the absolute values of the received sequence bit LLRs by $|\boldsymbol{l}|=(|l_1|, |l_2|, \cdots, |l_n|)$. They can be arranged in an ascending order via a permutation which we denote by $\boldsymbol\pi=(\pi_1, \ldots, \pi_n)$.
The main objective of a decoder is to estimate the noise effect ${\boldsymbol{z}}$ and correct the errors in the hard-decision received sequence to recover the transmitted codeword.
Depending on utilization of the LLR vector and the ordering permutation $\boldsymbol{\pi}$ we can distinguish several universal soft-decoding methods.

\subsection{Chase II Algorithm} \label{subsec:ChaseII}
The Chase II algorithm~\cite{chase1972class} begins by calculating the error syndrome of length $n-k$ via
\begin{equation} \label{eq:syndrome}
    \boldsymbol{s} = \boldsymbol{w} {\bf H}^\textrm{T},
\end{equation}
where $\bf{H}$ denotes a parity-check matrix of the code. In case $\boldsymbol{s}=\textbf{0}$, the algorithm outputs $\boldsymbol{w}$ as the decoded codeword. Otherwise, it proceeds by generating a set $\TC$ that contains $2^p$ test patterns, denoted as $\boldsymbol{t}$, by utilizing the $p$ least reliable positions of the LLR sequence $\pi_1,\ldots,\pi_p$. Each test pattern $\boldsymbol{t}$ in $\TC$ can have values of $0$ or $1$ in these positions, with $0$ elsewhere. 

The algorithm goes over the test patterns $\boldsymbol{t}\in\TC$. For each $\boldsymbol{y}'=\boldsymbol{w}\oplus\boldsymbol{t}$, bounded distance decoding is performed to obtain a codeword $\tilde{\boldsymbol{c}}$. From this decoded codeword, an error pattern ${\tilde{\boldsymbol{e}}}$ that relates $\boldsymbol{w}$ to $\tilde{\boldsymbol{c}}$ is determined by $\tilde{\boldsymbol{e}}=\boldsymbol{w}\oplus\tilde{\boldsymbol{c}}$. An analog weight is then assigned to each error pattern $\tilde{\boldsymbol{e}}$. If $\tilde{\boldsymbol{e}}$ has 1's in positions $j_1,\ldots,j_N$, the {\em analog weight} is computed as the sum of the absolute values of the respective LLRs via $\epsilon(\tilde{\boldsymbol{e}})=|l_{j_1}|+|l_{j_2}|+\cdots+|l_{j_N}|$. Subsequently, the error pattern $\tilde{\boldsymbol{e}}_{\mathrm{best}}$ with the smallest $\epsilon(\tilde{\boldsymbol{e}})$ among all error patterns is identified as the best estimate of the noise effect. Finally, error correction is performed to generate the estimated codeword $\boldsymbol{c}_{\mathrm{best}}$ as the output, using $\boldsymbol{c}_{\mathrm{best}}=\boldsymbol{w}\oplus\tilde{\boldsymbol{e}}_{\mathrm{best}}$.

\subsection{ORBGRAND Algorithm} \label{subsec:ORBGRAND}

The ORBGRAND algorithm~\cite{duffy2022ordered} starts with a pre-defined list of error patterns $\hat{\boldsymbol{e}}^{(1)},\hat{\boldsymbol{e}}^{(2)},\ldots,\hat{\boldsymbol{e}}^{(q)}, \ldots$, where $\hat{\boldsymbol{e}}^{(q)}=(\hat{e}_1^{(q)},\hat{e}_2^{(q)},\ldots,\hat{e}_n^{(q)})$, $\hat{e}_i^{(q)} \in \{0,1\}$, $i=1,2,\ldots,n$ sorted in the order of their approximate likelihood. 
To achieve this, a \textit{logistic weight} of a pattern $\hat{\boldsymbol{e}}^{(q)}$ that approximates its analogue weight is defined by 
\begin{equation} \label{eq:02}
    \omega_L(\hat{\boldsymbol{e}}^{(q)}) = \sum_{i=1}^{n} i\cdot \hat{e}_i^{(q)}.
\end{equation}
i. e., the sum of indices corresponding to non-zero elements in $\hat{\boldsymbol{e}}^{(q)}$. The patterns are generated in the order of their logistic weight. The first pattern $\hat{\boldsymbol{e}}^{(1)}$ satisfies $\omega_L(\hat{\boldsymbol{e}}^{(1)})=0$, and consists of all zeros. The second pattern $\hat{\boldsymbol{e}}^{(2)}$ with logistic wight one contains a single one in the least reliable position. It is followed by $\hat{\boldsymbol{e}}^{(3)}$ of logistic weight two that contains a single one in the next least reliable position and so on. 


The LLR-ordering permutation $\boldsymbol{\pi}$, obtained from the received sequence, is then employed to map the positions to 1's of the pre-defined patterns to the proper least reliable positions producing the error patterns $\Tilde{\boldsymbol{e}}^{(1)},\Tilde{\boldsymbol{e}}^{(2)},\ldots,\Tilde{\boldsymbol{e}}^{(q)}, \ldots$ that will be tested. The patterns are tested according to their order. A pattern is added to the hard-decision received sequence $\boldsymbol{w}$ to check if it leads to a codeword via $(\boldsymbol{w} \oplus \Tilde{\boldsymbol{e}}^{(Q)}) { \bf H}^\textrm{T} =  \boldsymbol{0}.$
This process continues until either a codebook member is found, or no codeword is found within the limit of $Q_{\mathrm{max}}$ codebook queries. In the latter scenario, the algorithm abandons further search and returns a decoding failure. The ORBGRAND algorithm is given by Algorithm \ref{alg:ORBGRAND}.

\begin{algorithm}
\caption{ORBGRAND Algorithm}\label{alg:ORBGRAND}
\begin{algorithmic}[1]
\Require {Received signal $\boldsymbol{y}$, codebook $\CC$, maximum number of error pattern queries $Q_{\mathrm{max}}$}.
\Ensure {decoded  codeword $\boldsymbol{c}_{\mathrm{best}}$}.
\State Sort the absolute LLR values $|\boldsymbol{l}|$ of the input $\boldsymbol{y}$, and determine the permutation vector $\boldsymbol\pi$.
\State $q=0$ \Comment{Initialize the number of queries.} 
\While{$\boldsymbol{w} \oplus \Tilde{\boldsymbol{e}}^{(q)} \notin \CC$ \textbf{and}  $q<Q_{\mathrm{max}}$}
    \State Generate the next most likely error pattern $\hat{\boldsymbol{e}}^{(q)}$ (or retrieve from a query list) and produce its LLR-ordered version $\Tilde{\boldsymbol{e}}^{(q)}$.
    \If{$\boldsymbol{w} \oplus \Tilde{\boldsymbol{e}}^{(q)} \in \CC$}
        \State \textbf{Return} $\boldsymbol{c}_{\mathrm{best}} = \boldsymbol{w} \oplus \Tilde{\boldsymbol{e}}^{(q)}$
    \Else
    \State $q=q+1$ 
    \EndIf
\EndWhile
\end{algorithmic}

\end{algorithm}

Equation~\ref{eq:02} approximates the ordered reliabilities by a line with zero intercept and unit slope, representing the basic version of ORBGRAND. This approximation adequately serves its purpose in lower SNRs when ordered reliabilities are near-linear. However, it becomes less effective in higher SNR scenarios where the ordered reliabilities deviate from linearity. To address this, multi-line variations of ORBGRAND have been introduced in~\cite{duffy2022ordered}, offering enhanced performance in higher SNRs.

\subsection{Proposed Ordered Reliability Direct Error-Pattern Testing (ORDEPT) Decoding Algorithm} \label{subsec:ORDEPT}

The algorithm operates with a set of PEPs selected in the order of their approximate likelihood. A PEP, denoted as $\boldsymbol{e}=(e_1,e_2,\ldots,e_n)$, is defined as a binary vector in which each element $e_i$, corresponding to a possible error occurring at the $i$-th position, is set to $1$. However, the last error position, which has the lowest likelihood and hence the highest index, remains $0$, along with all other elements of $\boldsymbol{e}$. The concept here is to determine if a potential PEP can be completed by an index for the last error position and lead to a valid codeword.

Applying a PEP denoted by $\boldsymbol{e}$ with $N-1$ ones in positions $e_{j_1}, e_{j_2}, \ldots, e_{j_{N-1}}$ to the hard-decision received sequence $\boldsymbol{w}$ corresponds to computation of a partial syndrome vector
\begin{equation} \label{eq:syndrome_prime}
        \Tilde{\boldsymbol{s}}=(\boldsymbol{w} \oplus {\boldsymbol{e}}) { \bf H}^\textrm{T}=\boldsymbol{s}+\boldsymbol{h}_{j_1}+ \boldsymbol{h}_{j_2} + \cdots+ \boldsymbol{h}_{j_{N-1}},
\end{equation} 
where $\boldsymbol{h}_{j_i}$ represents the $j_i$-th column of the code's parity check matrix $\bf{H}$, $i=1,\ldots,N-1$, and $\boldsymbol{s}$ is the $(n-k)$-bit syndrome vector that is calculated via \eqref{eq:syndrome}.

The partial syndrome $\Tilde{\boldsymbol{s}}$ can now be employed to identify the last error position, denoted as $j_*$, for which $\boldsymbol{w}\oplus\Tilde{\boldsymbol{e}}$ forms a valid codeword. In this context, $\Tilde{\boldsymbol{e}}$ represents the completed error pattern, where $\Tilde{e}_j=e_j$ for $j\neq j_{*}$, and $\Tilde{e}_{j_{*}}=1$. The process of finding $j_*$ is equivalent to locating a column within ${\bf H}$ such that $\Tilde{\boldsymbol{s}}=\boldsymbol{h}_{j_*}$. This mapping can be accomplished through a function that associates the decimal value of the column $\boldsymbol{h}_{j_*}$ of  $\bf{H}$ with its corresponding index $j_{*}$. This mapping is represented as a function $f(\Tilde{\boldsymbol{s}})$ defined as

    \begin{equation} \label{eq:analytic_function}
        f(\Tilde{\boldsymbol{s}})=\begin{cases}
        j_* & \mathrm{if}\quad \Tilde{\boldsymbol{s}}=\boldsymbol{h}_{j_*}\\
        -1 & \mathrm{otherwise,}
        \end{cases}.
    \end{equation}

Whereby, $f(\Tilde{\boldsymbol{s}})=-1$ signifies the absence of any column index $j_*$ such that $\boldsymbol{h}_{j_*}=\Tilde{\boldsymbol{s}}$. The function $f(\Tilde{\boldsymbol{s}})$ can be implemented with simple combinational logic involving $n-k$ binary inputs, depth of $\mathrm{log}_2(n-k)$, and $\log_2(n)$ outputs. In instances where $j_*$ is successfully identified and $f(\Tilde{\boldsymbol{s}})\neq -1$, the resulting codeword is computed by
\begin{equation} \label{eq:error_correction}
    \tilde{\boldsymbol{c}}=\boldsymbol{w}\oplus\Tilde{\boldsymbol{e}}.
\end{equation}

In the proposed algorithm a pre-defined list  $\hat{\boldsymbol{e}}^{(1)},\hat{\boldsymbol{e}}^{(2)},\ldots,\hat{\boldsymbol{e}}^{(q)}, \ldots$ of PEPs is pre-generated in the order of their likelihood. This can be achieved through methods like the build mountain routine and the landslide algorithm, as described in~\cite{duffy2022ordered}. The LLR-ordering permutation is applied to produce a list of candidate PEPs ${\boldsymbol{e}}^{(1)},{\boldsymbol{e}}^{(2)},\ldots,{\boldsymbol{e}}^{(q)}, \ldots$. The algorithm then proceeds to examine the list of candidate PEPs, either sequentially or in parallel, until it identifies the first instance, where $f(\Tilde{\boldsymbol{s}})\neq-1$. This signifies a scenario where a PEP can be completed to form an error-pattern $\Tilde{\boldsymbol{e}}$ that ultimately leads to a decoded codeword $\boldsymbol{c}_\textrm{best}=\tilde{\boldsymbol{c}}$ (see~\eqref{eq:error_correction}).

For extended codes, it is possible to reduce the number of PEPs that need to be queried. This is because the last bit of the syndrome effectively indicates whether the total number of errors is even or odd. In such situations, two separate lists of PEPs can be utilized: one for cases with an odd Hamming weight and another for cases with an even Hamming weight. This approach effectively reduces the number of PEPs examined for each codeword by a factor of two.

Furthermore, the performance of the proposed algorithm can be enhanced if it continues to search for additional candidate codewords after finding the first one. In such a scenario, the Euclidean distances between the channel soft output $\boldsymbol{y}$ and the modulated versions of all candidate codewords are calculated, and the candidate with the smallest distance is selected as the decoded codeword. Alternatively, the resulting codewords can be ranked based on the analog weights of the respective patterns. To maintain low complexity and minimize latency, a thresholding can be introduced, where the algorithm ceases its search when no new codeword is found within a certain number of queries, typically after $T$ queries following the most recent successful one. The thresholding is based on the observation that in case PEPs that lead to codewords form a cluster with close logistic weights, they have nearly equal chances to lead to the correct codeword. On the other had, if such cluster is followed by a few distant successful PEPs with very high logistic weight, these extra PEPs lead to the correct codeword very rarely.
The complete steps of the proposed algorithm (with thresholding) are detailed in Algorithm~\ref{alg:ORDEPT}.

\begin{algorithm}
\caption{ORDEPT Algorithm}\label{alg:ORDEPT}
\begin{algorithmic}[1]
\Require {Received signal $\boldsymbol{y}$, codebook $\CC$, maximum number of error pattern queries $Q_{\mathrm{max}}$}, PEP list, threshold $T$, $C_{\mathrm{max}}$ number of candidate codewords to be found.
\Ensure {Decoded codeword $\boldsymbol{c}_\mathrm{best}$.}
\If{$\boldsymbol{w}\in\CC$}
\State \textbf{return} $\boldsymbol{c}_\mathrm{best}=\boldsymbol{w}$
\Else
\State Sort the absolute LLR values $|\boldsymbol{l}|$ of the input $\boldsymbol{y}$, and determine the permutation vector $\boldsymbol\pi.$
\State $q=0$  \Comment{Initialize the index of the current PEP query.} 
\State $C=0$  \Comment{Initialize the number of codewords found by the algorithm.} 
\State $I_s=0$  \Comment{Initialize the index of the latest successful PEP query.} 
\State Calculate the syndrome via \eqref{eq:syndrome}.
\While{$C<C_{\mathrm{max}}$ \textbf{and}  $q<Q_{\mathrm{max}}$} 
    \State Retrieve the next PEP $\hat{\boldsymbol{e}}^{(q)}$ from the PEP list and apply $\boldsymbol{\pi}$ to acquire ${\boldsymbol{e}}^{(q)}$. Obtain $\Tilde{\boldsymbol{s}}$ via \eqref{eq:syndrome_prime}.
    \If{$f(\Tilde{\boldsymbol{s}})\neq-1$} 
     \State    Complete PEP by setting $\tilde{{e}}_j=e_j$ for $j \neq j_{*}$ and $\tilde{e}_{j*}=1$. Compute $\tilde{\boldsymbol{c}}$ using \eqref{eq:error_correction} and store it in $\LC$.
     \State $C=C+1$, \quad $I_s=q$, \quad $q=q+1$
    \Else
    \State $q=q+1$ 
    \EndIf
    \If{$C\neq0$ \textbf{and} $q>I_s+T$}
  \State    \textbf{break}
    \EndIf
\EndWhile
\State \textbf{return} $\boldsymbol{c}_\mathrm{best}=\argmin_{\tilde{\boldsymbol{c}}\in\LC} |\boldsymbol{y}-\tilde{\boldsymbol{x}}|^{2}$, where $\tilde{\boldsymbol{x}}=\tilde{\boldsymbol{x}}(\tilde{\boldsymbol{c}})$ is a BPSK modulated version of $\tilde{\boldsymbol{c}}$.
\EndIf
\end{algorithmic}
\end{algorithm}

The algorithm's simplicity primarily arises from its ability to directly identify the last error position within the noise effect using a simple analytic function. This approach aligns with the error pattern processing techniques employed in the DEPT algorithm, as described in~\cite{truhachev2020efficient}. In terms of hardware implementation, for each pattern both ORBGRAND and ORDEPT necessitate a syndrome computation logic based on adding the columns of the parity-check matrix corresponding to the error positions. Implementation of a combinational logic which outputs the bits of the column index equal to the computed syndrome would need a comparable or even smaller amount of resources, compared to the syndrome computation.

In the context of hardware implementations for list-based query algorithms like ORBGRAND, it is a common practice to simultaneously examine multiple PEPs, as highlighted in~\cite{condo2022iterative}. In our approach, we partition the entire PEP list into sets of queries, each containing $Q_{\textrm{s}}$ queries, and refer to each of these sets as shots. All patterns within a single shot are subjected to parallel testing. In the ORDEPT algorithm the shots are tested until either the required number of potential codewords, denoted by $C_{\mathrm{max}}$, is attained, or until a maximum of $T$ queries following the most recent successful PEP is reached.

The ORDEPT algorithm, as presented, can be viewed as a trade-off between Chase-like algorithms such as Chase II and DEPT, and the GRAND-type algorithms found within the ORBGRAND ``family'', effectively harnessing the strengths of both approaches. Chase II, known to be effective for a sufficiently large number of bit-flip positions $p$ and respective $2^p$ test patterns, offers powerful error correction capabilities. However, they come at the cost of increased complexity and latency due to multiple hard-decision decoding attempts and the need for a large number of soft reliability computations. On the other hand, the DEPT algorithm reduces the complexity by focusing on PEPs that do not necessitate full hard-decision decoding, with careful PEP selection. Nevertheless, it relies on a large number of PEPs and compares the reliability of a large number of resulting codewords. The ORDEPT algorithm that we propose in this work streamlines the decoding effort by utilizing PEPs organized in terms of approximate reliability, while maintaining the same block error probability (BLER) performance.

Whereas ORBGRAND similarly employs ordered reliability and a straightforward process for pattern testing, ORDEPT is more efficient. This is achieved by employing a significantly reduced number of patterns to swiftly identify the correct codeword. Moreover, the testing effort for a PEP in ORDEPT only marginally surpasses the testing complexity of a full error pattern in ORBGRAND. Consequently, ORDEPT offers diminished complexity and latency, making it a more suitable choice for product code decoding compared to ORBGRAND. This distinction arises from the capacity of ORDEPT to identify multiple potential codewords effectively, particularly with moderate list sizes, as demonstrated in Section~\ref{sec:SimulationResults}.

\subsection{ORDEPT for BCH Codes and Generalized ORDEPT} 
\label{subsec:ORDEPTBCH}

For BCH codes, which are specifically important due to their use in various concatenated coding schemes, the ORDEPT algorithm can be further simplified and generalized.

Consider a primitive $t$-error correcting extended binary $(2^r,2^r-tr-1)$ BCH code. For example the $(256,239)$ extended BCH code has $t=2$, whereas $r=8$ and the overall code length equals $n=2^r=256$. 
A BCH code is constructed from the elements of a Galois field $GF(2^r) = \{0,1,\gamma,\gamma^2,\cdots,\gamma^{2^r-2}\}$, where $\gamma$ is the primitive element and $p(\gamma)$ is the primitive polynomial of degree $r$. A parity-check matrix of the code can be constructed in the form 
\begin{equation}
\bf{H} = \begin{bmatrix}
\mathcal{B}(\gamma^0) & \mathcal{B}(\gamma^1) & \ldots & \mathcal{B}(\gamma^{2^{r}-2})  & 0\\
\mathcal{B}(\gamma^0) & \mathcal{B}(\gamma^3) & \ldots & \mathcal{B}(\gamma^{3(2^{r}-2)})  & 0\\
\vdots & \vdots &  & \vdots & \vdots \\
\mathcal{B}(\gamma^0) & \mathcal{B}(\gamma^{2t-1}) & \ldots & \mathcal{B}(\gamma^{(2t-1)(2^{r}-2)})  & 0\\
1 & 1 & \cdots & 1 & 0  
\end{bmatrix}\hspace{-1mm}, 
\label{eq:HBCH}
\end{equation}
where $\mathcal{B}(\cdot)$ denotes an $r$-bit binary column vector of a polynomial coefficient representation of a GF element. 
The last row of the parity-check matrix contains 1's and is used to detect the parity of the number of errors. 

Consider the partial syndrome vector $\Tilde{\boldsymbol{s}}$ in  (\ref{eq:syndrome_prime}). Let us denote the first $r$ bits of the partial syndrome vector $\Tilde{\boldsymbol{s}}$ by $\Tilde{\boldsymbol{s}}_1$. The ORDEPT attempts to find a column index $j_*$ such that $\Tilde{\boldsymbol{s}}=\boldsymbol{h}_{j_*}$. Let us define the sub-matrix ${\bf H}_1$ of the parity-check matrix (\ref{eq:HBCH}) constructed from the first $r$ rows. The set of columns $\boldsymbol{h}^{(1)}_{1},\cdots,\boldsymbol{h}^{(1)}_{n}$ of ${\bf H}_1$ coincides with a set of all $2^r$ possible distinct binary vectors of length $r$. On the other hand, there exist $2^r$ distinct column indices.
Hence, ${j_*}$ such that $\Tilde{\boldsymbol{s}}_1=\boldsymbol{h}^{(1)}_{j_*}$ can be uniquely determined and expressed as a permutation $j_*=\tilde{\pi}(\eta(\Tilde{\boldsymbol{s}}_1))$, where $\eta(\cdot)$ denotes a decimal number representation of a binary vector. Once the unique candidate $j_*$ is determined based on the first $r$ parity-check rows of $\bf{H}$, we can then check if $\Tilde{\boldsymbol{s}}=\boldsymbol{h}_{j_*}$ is satisfied for the remaining $(t-1)r+1$ bits positions as well, via a simple `and' gate logic function.

While the ORDEPT algorithm presented in Section~\ref{subsec:ORDEPT} is trying to complete each PEP with one error position, we can further generalize it to ORDEPT$x$, $x>1$, that tries to complete each PEP with upto $x$ remaining error positions, where $x < t$. For $x=0$ ORDEPT$x$ coincides with ORBGRAND, while for $x=1$ we obtain the regular ORDEPT described in Section~\ref{subsec:ORDEPT}.

For the case of primitive extended BCH codes defined by (\ref{eq:HBCH}), completion of a PEP with $x$ remaining error positions can be accomplished via a hard decision decoding of an $x$-error correcting extended BCH code defined by a parity-check matrix which consists of the first $xr$ rows of $\bf{H}$ appended by a row of $1$s.

Note that for $x \leq 3$ BCH decoding can be performed with help of a few look up tables (that can alternatively be implemented as combinational logic functions) of length $2^r$ (see~\cite{smith2011staircase}), while for $t>3$ a multi-step algorithm such as Berlekamp-Massey, Peterson, etc. would be required \cite{LinCostello}. Therefore, for $t>3$ ORDEPT$x$, where $x =1,2$, or $3$, can provide a lower latency and complexity approach in comparison with Chase II algorithm or ORBGRAND that requires a very large number of patterns (queries) to be tested.

\section{ORDEPT as Component Decoder} 
\label{sec:Component}

In this section, the ability of the proposed ORDEPT decoder to serve as a component decoder for iteratively decodable product-like codes is demonstrated. We consider two types of codes: the original block product codes~\cite{elias1954error} and the oFEC code designed for single wavelength transmission over optic fiber~\cite{oFEC}. The oFEC is a representative convolutional product-like code which belongs to the families of braided block codes~\cite{truhachev2003braided,feltstrom2009braided} and zipper codes~\cite{sukmadji2019zipper}.

\subsection{Decoding Product Codes with ORDEPT} 
\label{sec:DecodeProduct}

Product codes are constructed by concatenation of two types of block codes denoted by $\CC_i$, $i=1,2$, with parameters $(n_i,k_i)$ where  $n_i$ represents the code length and $k_i$ is the code's dimension~\cite{elias1954error}. To create a product code, the information bits are organized into a rectangular array with dimensions $k_2\times k_1$. Subsequently, each row of this array is expanded to length $n_1$ using the parity-check bits derived from encoding of the $k_1$ information bits of that row with $\CC_1$, while each column is extended to length $n_2$ using the parity-check bits produced by encoding of the column with $\CC_2$.

For the soft iterative decoding of product codes, it is essential to obtain soft output to update the reliability of each bit. Typically, Pyndiah's decoding method, as detailed in \cite{pyndiah1998near}, is employed for this purpose. During each iteration, rows and columns are decoded in two alternating half iterations. As shown in \cite{pyndiah1998near}, to generate soft output, we require at least one less likely competitor codeword, denoted as $\tilde{\boldsymbol{c}}$, from the decoder of each component code. These competitor codewords help to compute soft output, i.e., LLRs of the bits that can be exchanged between half iterations, transitioning from row decoding to column decoding and vice versa. We represent the soft output of the $(j-1)$-th half iteration by $\hat{\boldsymbol{y}}^{j-1}$, which subsequently becomes the input for the $j$-th half iteration. With a list of candidate codewords, denoted as $\LC$, already available from the proposed ORDEPT decoder (as described in Section~ \ref{subsec:ORDEPT}), the soft output can be computed for the $i$-th bit using Pyndiah's method~\cite{pyndiah1998near}. Initially, we select $\boldsymbol{c}_{\mathrm{best}}$ as the decoded codeword, determined via $\boldsymbol{c}_{\mathrm{best}}=\argmin_{\tilde{\boldsymbol{c}}\in\LC} |\hat{\boldsymbol{y}}^{j-1}-\tilde{\boldsymbol{x}}|^{2}$, where $\tilde{\boldsymbol{x}}=\hat{\boldsymbol{x}}(\tilde{\boldsymbol{c}})$ represents a BPSK modulated version of $\tilde{\boldsymbol{c}}$. Subsequently, if there exists a codeword $\boldsymbol{c}^{*}$ within the list $\LC$ that differs from $\boldsymbol{c}_{\mathrm{best}}$ in terms of the $i$-th bit value, where $\boldsymbol{c}^{*}=\argmin_{{\tilde{\boldsymbol{c}}\in\mathcal{L}:\tilde{c}_{i}\neq{c}_{{\mathrm{best}}_i}}} |{\hat{\boldsymbol{y}}^{j-1}}-\tilde{\boldsymbol{x}}|^{2}$, then the initial soft output for the $i$-th bit is determined by 
\begin{equation} \label{eq:Pyndiah_Update}
    \hat{y}^j_{i}=\frac{x_{{\mathrm{best}}_i}(|\hat{\boldsymbol{y}}^{j-1}-\boldsymbol{x}^{*}|^{2}-|\hat{\boldsymbol{y}}^{j-1}-{\boldsymbol{x}}_{\mathrm{best}}|^2)}{4},
\end{equation}
where $x_{{\mathrm{best}}_i}$ is the modulated representation of the $i$-th decoded codeword bit, i.e., $x_{\mathrm{best}_i}\in \{-1,1\}$ and $\boldsymbol{x}^*$ denotes the modulated form of $\boldsymbol{c}^*$. In the case that the candidate codewords within the list $\LC$ unanimously agree on the value of the $i$-th bit the expression

\begin{equation}  \label{eq:Beta}
    \hat{y}^j_{i}=\beta^{(j)}\times x_{{\mathrm{best}}_i}
\end{equation}
is used as the initial soft output for some constant $\beta^{(j)}\geq0$. The extrinsic information of each bit is calculated via
    \begin{equation} \label{eq:extrinsic}
         \omega_i = \hat{y}^{j-1}_{i} - y_i.
    \end{equation} 

The ultimate updated soft information for each bit is derived by incorporating a scaled extrinsic information value into the original channel information through
    \begin{equation} \label{eq:07}
        \hat{y}^j_{i}=y_i+\alpha^{(j)}\omega_i,
    \end{equation}
for some constant value of $\alpha^{(j)}$, $0<\alpha^{(j)}\leq1$. The steps of the ORDEPT-based Pyndiah's decoding algorithm with a general candidate codeword list $\LC$ are given in the Algorithm~ \ref{alg:Pyndiah}.

\begin{algorithm}
\caption{Chase-Pyndiah's Decoding Algorithm}\label{alg:Pyndiah}
\begin{algorithmic}[1]
\Require {Received signal $\boldsymbol{y}$, list of candidate codewords $\LC$, soft output of the previous iteration $\hat{\boldsymbol{y}}^{j-1}$, $\alpha^{(j)}$, $\beta^{(j)}$. }
\Ensure {Soft output $\hat{\boldsymbol{y}}^{j}$. }

\State Set $\boldsymbol{c}_{\mathrm{best}}=\argmin_{\tilde{\boldsymbol{c}}\in\LC} |\hat{\boldsymbol{y}}^{j-1}-\tilde{\boldsymbol{x}}|^{2}$ as the decoded codeword.
\For{$i=1:n$}
   \If{$c_{\mathrm{best}_i}=\tilde{c}_i$ for all $\tilde{\boldsymbol{c}}\in\LC$}
\State $\hat{y}^j_{i}=\beta^{(j)}\times x_{{\mathrm{best}}_i}$
\Else 
\State{ Compute $\boldsymbol{c}^{*}=\argmin_{\{\tilde{\boldsymbol{c}}\in\mathcal{L}:\tilde{c}_{i}\neq{c}_{{\mathrm{best}}_i}\}} |{\hat{\boldsymbol{y}}^{j-1}}-\tilde{\boldsymbol{x}}|^{2}$ }.
\State         $\hat{y}^j_{i}=\frac{x_{\mathrm{best}_i}(|\hat{\boldsymbol{y}}^{j-1}-\boldsymbol{x}^{*}|^{2}-|\hat{\boldsymbol{y}}^{j-1}-{\boldsymbol{x}}_{\mathrm{best}}|^{2})}{4}$
\EndIf
\State Calculate extrinsic information via $\omega_i = \hat{y}^{j-1}_{i} - y_i$,
\State $\hat{y}^j_{i}=y_i+\alpha^{(j)}\omega_i$
\EndFor

\State \textbf{return} $\hat{\boldsymbol{y}}^{j}$
\end{algorithmic}
\end{algorithm}


\subsection{Adaptive Hybrid SISO Decoding Algorithm} 

\label{sec:AdaptivePyndiah}
In the conventional Chase-Pyndiah algorithm, the $\alpha$ and $\beta$ remain constant for all component codes in each half iteration. In a recent work, an adaptive algorithm is proposed for Chase-Pyndiah method based on the analogue weight $\epsilon(\boldsymbol{w}+\boldsymbol{c}_\mathrm{best})$ (see Section~\ref{subsec:ChaseII}) of the error pattern that leads to the decoded codeword $\boldsymbol{c}_{\mathrm{best}}$ \cite{deng2023adaptive}. Similarly to the Eucledian distance, the analogue weight can be used as an indicator of a codeword's reliability and applied to update soft information. In \cite{deng2023adaptive}, during $j$-th half iteration, upper limits $a_{\alpha}^{(j)}$ and $a_{\beta}^{(j)}$ for $\alpha$ and $\beta$, lower limits $b_{\alpha}^{(j)}$ and $b_{\beta}^{(j)}$ for $\alpha$ and $\beta$ and decreasing coefficients $k_{\alpha}^{(j)}$ and $k_{\beta}^{(j)}$ for $\alpha$ and $\beta$ are defined. Then, for each component codeword in each half-iteration, $\alpha$ and $\beta$ can fluctuate between the upper and lower limits as functions of the analogue weight.

The proposed ORDEPT decoder naturally provides us with access to an alternative metric for evaluating the reliability of the codewords. Given that the PEPs in ORDEPT are sorted in decreasing order of likelihood, the identified candidate codewords in the decoder's list are consequently sorted from most likely to least likely (approximately). Therefore, the codeword's position in the list indicates its reliability with lower indices denoting higher reliability. 

Therefore, we propose an adaptive decoder, where the decision making process is performed based on both the traditional distance-based and the new order-based metrics, with a weighting factor of $0.5$ for each metric. 
The details of the proposed adaptive hybrid method of soft information factor calculation are given in Algorithm \ref{alg:Adaptive_Pyndiah}. The $\alpha^{(j)}$ and $\beta^{(j)}$ obtained at the output of Algorithm \ref{alg:Adaptive_Pyndiah} can be used directly as inputs in Algorithm \ref{alg:Pyndiah}.

\begin{algorithm}
\caption{Adaptive Hybrid Algorithm of Factor calculation}\label{alg:Adaptive_Pyndiah}
\begin{algorithmic}[1]
\Require {$\epsilon(\Tilde{\boldsymbol{e}}_\mathrm{best})$, $a_{\alpha}^{(j)}$ , $a_{\beta}^{(j)}$, $b_{\alpha}^{(j)}$, $b_{\beta}^{(j)}$, $k_{\alpha}^{(j)}$, $k_{\beta}^{(j)}$, index of $\boldsymbol{c}_\mathrm{best}$ in the ORDEPT decoder list denoted by $i_\textrm{best}$.}
\Ensure {$\alpha^{(j)}$, $\beta^{(j)}$ }

\State $\alpha^{(j)}=a_{\alpha}^{(j)}-k_{\alpha}^{(j)}\times(0.5\times \epsilon(\Tilde{\boldsymbol{e}}_\mathrm{best})+0.5\times\frac{i_\textrm{best}-1}{2})$
\If{$\alpha^{(j)}<b_{\alpha}^{(j)}$}
\State $\alpha=b_{\alpha}^{(j)}$
\EndIf
\State $\beta^{(j)}=a_{\beta}^{(j)}-k_{\beta}^{(j)}\times(0.5\times \epsilon(\Tilde{\boldsymbol{e}}_\mathrm{best})+0.5\times\frac{i_\textrm{best}-1}{2})$
\If{$\beta^{(j)}<b_{\beta}^{(j)}$}
\State $\beta^{(j)}=b_{\beta}^{(j)}$
\EndIf
\end{algorithmic}
\end{algorithm}

\subsection{Sliding Window ORDEPT for oFEC} 
\label{sec:oFEC}

The oFEC is a recently introduced FEC for optical transport network (OTN), standardized by the Open ROADM specification document~\cite{oFEC}, and designed for high-throughput fiber-optical communications. The oFEC structure is capable of providing a net coding gain (NCG) of 11.1 dB for BPSK or quadrature phase shift keying (QPSK), and 11.6 dB for quadrature amplitude modulation (QAM) with $16$ constellation points (16QAM) on AWGN channel.  The code necessities a pre-FEC BER threshold of $2\times10^{-2}$ to deliver BER below $10^{-15}$ after three soft and two hard decision decoding iterations~\cite{wang2022investigation}. It has also been proposed for the next generation standard that operates at $800$~Gbps~\cite{wang2022investigation}.


\begin{figure*}[t]
    \centering
    \includegraphics[scale=0.16]{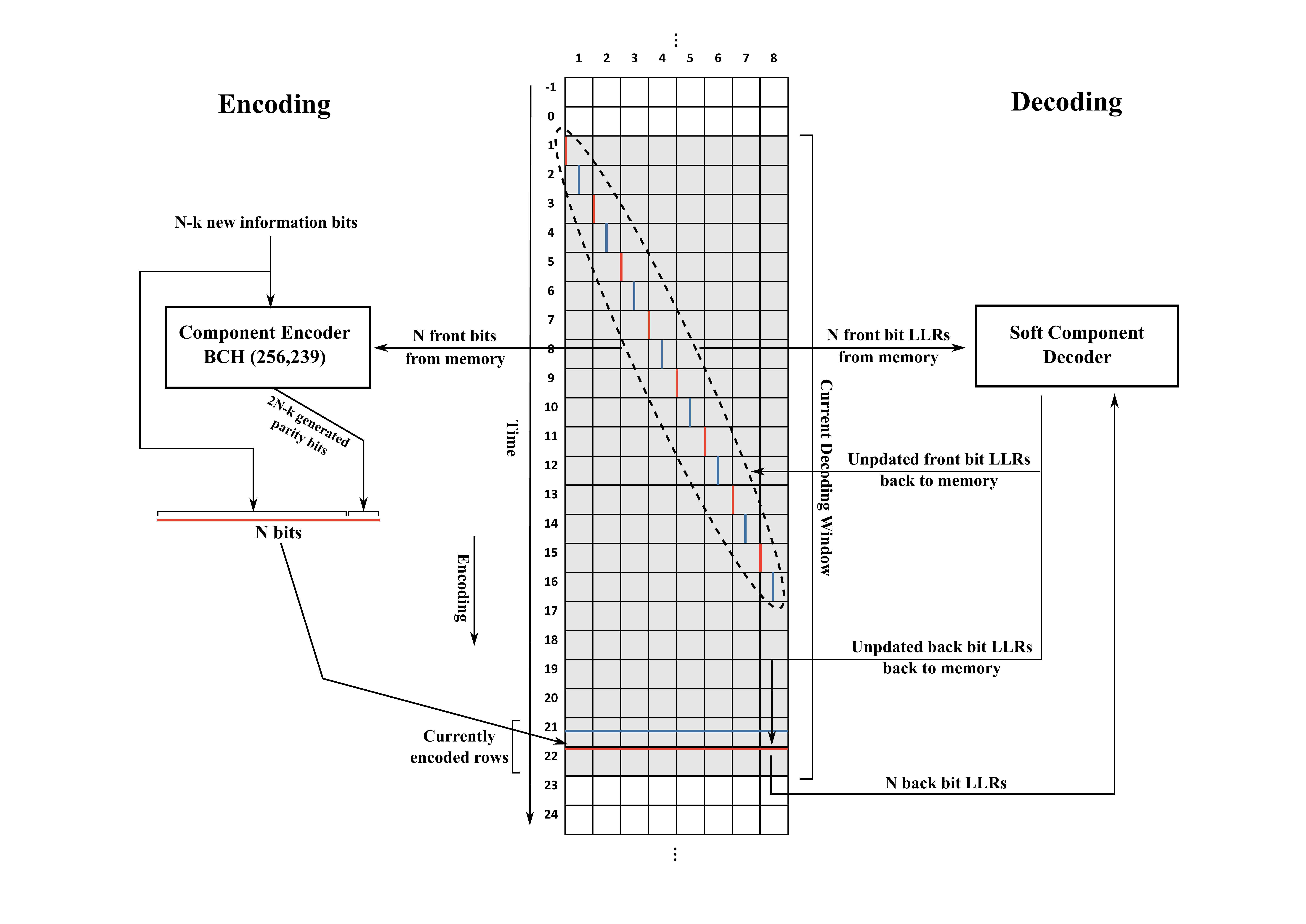}
    \caption{oFEC code array, encoding and decoding.}
    \label{fig:oFEC}
\end{figure*}

The oFEC, a convolutional product code type FEC, is defined by means of an infinite data array depicted in Fig.~\ref{fig:oFEC}. Each row of the array contains $N=128$ encoded data bits. To facilitate the encoding process, the array is sub-divided into $B\times B$ squared blocks where $B=16$. The extended BCH(256,239) codes serve as the component codes in the oFEC. Each codeword of length $2N=256$ bits (and dimension $k=239$) is split into $N$ back bits and $N$ front bits. The locations of the bits of one component codeword are shown by the red lines in Fig.~\ref{fig:oFEC}. The eight vertical red lines show $N/B$ groups of $B$ bits which form the back bits of the codeword. The red horizontal line shows the front bits of the same codeword.

The encoding process is depicted on the left side of Fig.~\ref{fig:oFEC}. To encode one component codeword, the $N$ front bits (shown by the vertical red lines) are read from the memory. They are combined with the $k-N=111$ input information bits and encoded by the systematic component encoder. The information bits, together with the computed $2N-k=17$ parity-check bits are permuted (see~\cite{oFEC}) and placed into the current encoded row (shown by the horizontal red line) of the oFEC array. We can see that each bit in the oFEC array belongs to two distinct component codewords. In the first codeword it plays a role of a back bit and in the second it is one of the front bits. Two rows of $B\times B$ blocks ($2B=32$ codewords) are encoded in parallel at each time as the encoding process moves down the array. The encoded data is modulated and transmitted over the channel.

At the receiver side (depicted on the right side of Fig.~\ref{fig:oFEC}) the soft demodulator obtains the intrinsic LLRs of each bit. These may be stored in an oFEC array at the receiver end, together with the extrinsic LLRs updated throughout the decoding iterations. 
The decoding iterations are organized in a sliding-window fashion. A window of $3+N/B=11$ rows of $B\times B$ blocks is dedicated to each iteration. Each iteration processes $2B$ component codes in parallel (shown by the area shaded in gray within the ``current iteration window'' in  Fig.~\ref{fig:oFEC}). Soft component decoders receive the intrinsic and extrinsic LLRs and produce the updated extrinsic LLRs at the output. Following that, all windows are shifted by two block rows.

Since the oFEC can be regarded as a convolutional type product code, the component codes are typically decoded and exchange soft information via the Chase-Pyndiah algorithm~\cite{pyndiah1998near,5709859,1306614,deng2023adaptive}. The required NCG may be achieved by three soft followed by two hard decision decoding iterations~\cite{wang2022investigation}.

\begin{table}[t]
       \caption{Equivalent uncoded BER and instantaneous SNR of the front and back bits for the first three iterations of soft decoding of the BPSK-modulated oFEC transmitted over AWGN channel with $E_b/N_0 = 3.89$ dB.}
    \centering
    \begin{tabular}{ccccc}
        \toprule
        \textbf{Iteration} & \multicolumn{2}{c}{\textbf{BER}} & \multicolumn{2}{c}{\textbf{SNR (dB)}}\\
         \cmidrule(lr){2-3}
         \cmidrule(lr){4-5}
        {} & Front & Back & Front & Back\\
        \midrule
        $1^{st}$ Iteration & $1.967\times10^{-2}$ & $1.417\times10^{-2}$ & $3.888$ & $4.428$ \\
        $2^{nd}$ Iteration & $1.131\times10^{-2}$ & $2.127\times10^{-3}$ & $4.766$ & $6.732$\\
        $3^{rd}$ Iteration & $6.198\times10^{-4}$ & $1.277\times10^{-5}$ & $7.792$ & $10.094$\\
        \bottomrule
    \end{tabular}
    \label{tab:FrontBackUncodedBER}
\end{table}

The proposed ORDEPT algorithm is applied to the oFEC following Algorithm~\ref{alg:Pyndiah}. There is, however, an important difference. Front and back bits of each component codeword are experiencing different levels of noise and interference since they are positioned in two distinct memory areas in the oFEC structure. The front bits are passed by sliding the current iteration window first, followed by the back bits. Table~\ref{tab:FrontBackUncodedBER} displays the BER of the front and back bits for the first, second, and third soft decoding iterations for $E_b/N_0=3.89$ dB (the operating point with the pre-FEC threshold of $2\times10^{-2}$). It reveals that the instantaneous SNR of the front bits is always greater than that of the back bits, while the equivalent uncoded BER is always smaller. This implies that it is impractical to employ the same weight and reliability factors $\alpha$ and $\beta$ for all the bits.  

According to the structure of the oFEC, we propose to employ two series of values of $\alpha$ and $\beta$ in each iteration for the front and back bits, respectively. While the oFEC system is decoded with $I$ soft iterations, $2I$ values for each parameter $\alpha$ and $\beta$ are utilized, where $\alpha^{(2j-1)}$ and $\beta^{(2j-1)}$ are used for the front bits, while the back bits are decoded using $\alpha^{(2j)}$ and $\beta^{(2j)}$, $j = 1,2,\ldots,I$.

\section{Numerical Results} \label{sec:SimulationResults}

This section presents a series of simulation results that illustrate the improvements achieved by ORDEPT in terms of error rates and decoding complexity compared to the state-of-the-art approaches, both in individual decoding as well as iterative decoding of product-like codes. All simulations are conducted assuming transmission over the AWGN channel with BPSK signaling. Following the error rate and complexity assessments, we present a comparison of the estimated required number of clock cycles which is an indicator of latency in hardware implementations. 

\subsection{Simulation Results}


\begin{figure}[!t]
\setlength{\unitlength}{1mm}
   \begin{picture}(0,75)(0,0)
   \put(0,0){\includegraphics[scale=0.127]{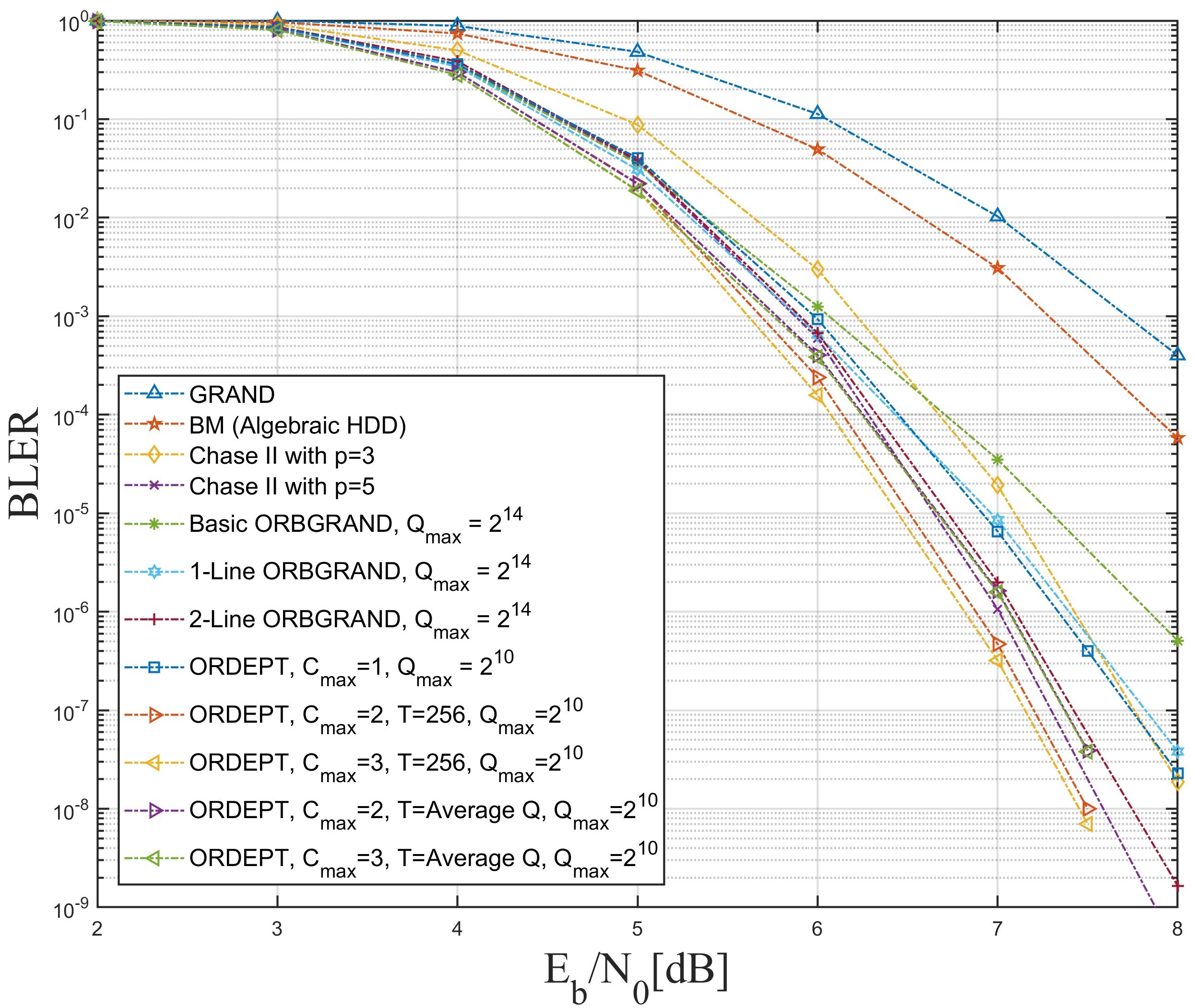}}
\end{picture}
\caption{Performance evaluation of ORDEPT in comparison with existing decoders for $\mathrm{BCH}(256,239)$.}
   \label{Fig:BLERBCH256239}
\end{figure}

Fig. \ref{Fig:BLERBCH256239} illustrates the BLER performance of ORDEPT compared to various existing decoding methods for the extended $\mathrm{BCH}(256,239)$ code. These methods encompass both hard input decoders like GRAND and Berlekamp-Massey (BM), and soft input decoders like the Chase II algorithm, which systematically flips all combinations of $p$ least reliable bit positions. Additionally, we include the basic and multi-line versions of ORBGRAND. In our evaluations of ORDEPT and ORBGRAND variants, we carefully select the smallest $Q_{\mathrm{max}}$ such that a significant performance loss would occur in case $Q_{\mathrm{max}}$ is further reduced. While the performance of all the algorithms improves with the increase of $Q_\mathrm{max}$, we selected $Q_\mathrm{max} = 2^{10}$ for ORDEPT and $Q_\mathrm{max} = 2^{14}$ for ORBGRAND variants as two representative values for a fair comparison, where ORDEPT improves both performance and latency. The SNR values required for ORDEPT with $C_\mathrm{max}=3$ and 2-line ORBGRAND to achieve BLER equal to $10^{-5}$ for several $Q_\mathrm{max}$ are given in Table~\ref{tab:Qmax}. For the last two ORDEPT simulations, we set the threshold parameter $T$ equal to the average number of queries in ORBGRAND variants. The results demonstrate that ORDEPT with $C_{\mathrm{max}}=1$ exhibits BLER performance which is similar to 1-line ORBGRAND, while ORDEPT with $C_{\mathrm{max}}=3$, $T=256$ shows approximately $0.2$~dB improvement over 2-line ORBGRAND at lower BLER. 

\begin{table}[t]
       \caption{SNR required to reach BLER equal to $10^{-5}$.}
    \centering
    \begin{tabular}{cccccc}
        \toprule
         & \multicolumn{2}{l}{\textbf{ORDEPT}, $\bf{C}_{\mathrm{\bf{max}}}=3$} & \multicolumn{3}{c}{\textbf{2-line ORBGRAND}}\\
         \cmidrule(lr){2-3}
         \cmidrule(lr){4-6}
        $\bf{Q}_{\mathrm{\bf{max}}}$ & $2^{10}$ & $2^{9}$ & $2^{14}$ & $2^{13}$ & $2^{9}$  \\
        \midrule
        SNR [dB] & $6.44$ & $6.47$ & $6.67$ & $6.76$ & $7.58$ \\
        \bottomrule
    \end{tabular}
    \label{tab:Qmax}
\end{table}

While the performance of all the algorithms improves with $Q_{\mathrm{max}}$, we selected $Q_{\mathrm{max}}=2^{10}$ for ORDEPT and $Q_{\mathrm{max}}=2^{14}$ for ORBGRAND variants as two representative values for a fair comparison, where ORDEPT improves performance, complexity and latency (see discussion below). The SNR values required for ORDEPT and 2-line ORBGRAND to achieve BLER level of $10^{-5}$ for several $Q_{\mathrm{max}}$ are given in Table 1.

It is noteworthy that the maximum number of queries, denoted as $Q_{\mathrm{max}}$ required to identify a codeword, is $16$ times smaller for the proposed ORDEPT algorithm compared to the ORBGRAND variants. Besides the reduction of the latency, discussed in more detail further, smaller $Q_{\mathrm{max}}$ also translates into a smaller number of sorted LLRs required in ORDEPT compared to ORBGRAND. Our simulations indicate that in the worst-case scenario, ORBGRAND necessitates identification and sorting of $45$ smallest Log-Likelihood Ratios (LLRs), while ORDEPT requires only $31$.



\begin{figure}[!t]
\setlength{\unitlength}{1mm}
   \begin{picture}(0,75)(0,0)
   \put(0,0){\includegraphics[scale=0.142]{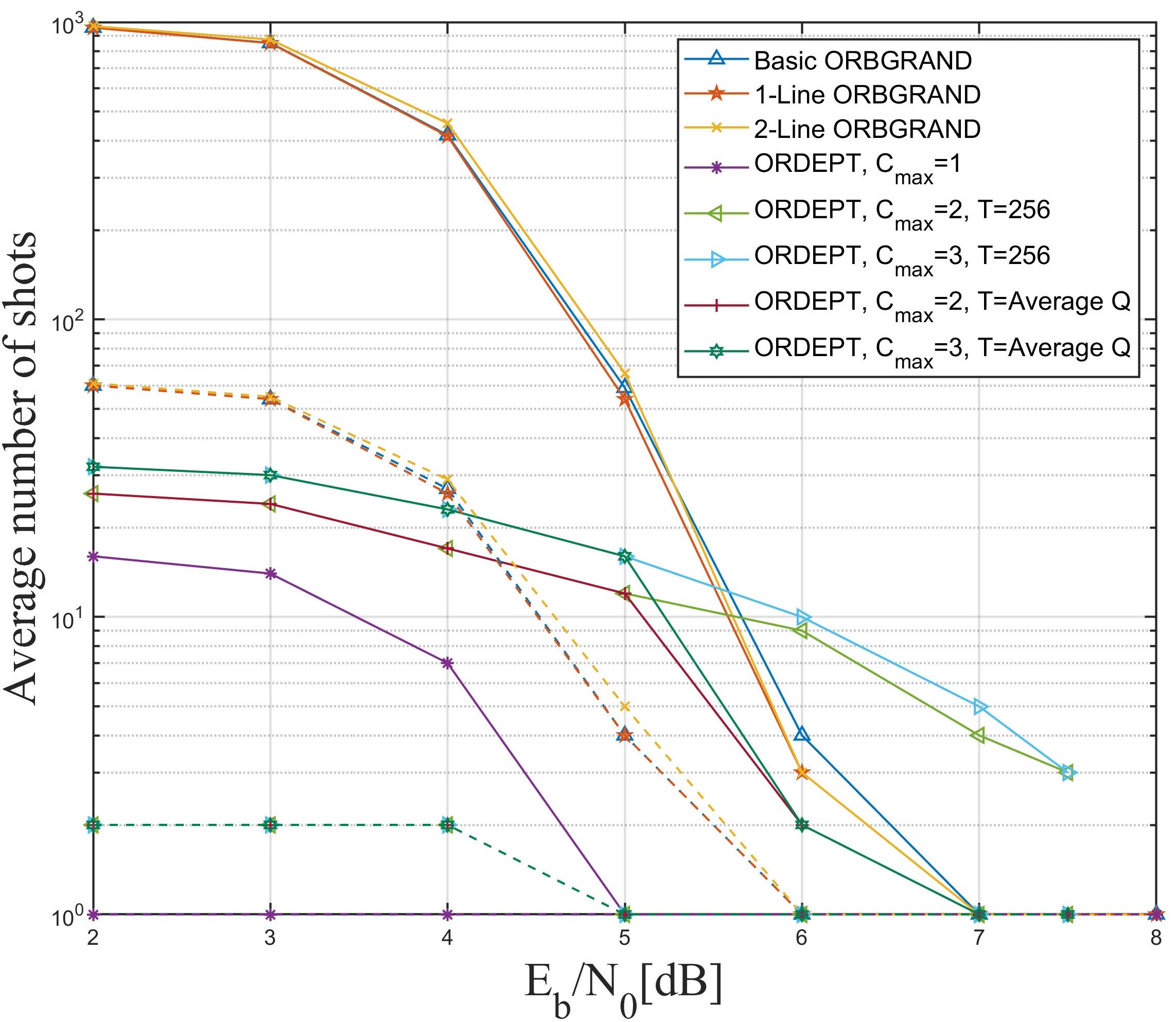}}
\end{picture}
\caption{The average number of shots to decode a codeword with shot size 16 (solid lines) and shot size 256 (dashed lines) using ORDEPT and ORBGRAND variants for $\mathrm{BCH}(256,239)$.}
   \label{Fig:ComplexityBCH256239}
\end{figure}


To compare the complexity of ORDEPT and various ORBGRAND variants, we examine the average number of shots required to decode a received sequence. Here, a shot refers to a group of queries processed concurrently which is a common practice in high-speed GRAND implementations \cite{condo2022fixed}. The complexity then translates into the circuit power dissipation.

The results are depicted in Fig.~\ref{Fig:ComplexityBCH256239}.  
We investigate two shot sizes, specifically $Q_{\textrm{s}}=16$ (represented by solid lines) and $Q_{\textrm{s}}=256$ (indicated by dashed lines). The results reveal that when $Q_{\textrm{s}}=256$, the proposed ORDEPT algorithm requires, in average, at most two shots across all SNRs, whereas ORBGRAND demands at least four shots for SNRs below $5$~dB. The advantage of ORDEPT in terms of the number of shots becomes even more pronounced when $Q_{\textrm{s}}=16$ is considered. By considering both Fig.~\ref{Fig:BLERBCH256239} (which illustrates BLER) and Fig.~\ref{Fig:ComplexityBCH256239} (depicting latency), we can see that ORDEPT, when configured with $T$=average~$Q$, outperforms ORBGRAND variants in terms of both the performance and latency.

The number of clock cycles that are needed to decode a codeword in a hardware implementation is the ultimate measure of the decoding latency. In a recent work, where a fixed latency architecture is proposed for ORBGRAND~\cite{condo2022fixed}, the required number of clock cycles for decoding of a codeword is given by
\begin{equation} \label{eq:ClockCycle}
    L = Q_{\mathrm{max}}/Q_{\mathrm{s}} + 2 + \mathrm{log}_2n \quad \quad \quad \quad
\end{equation}
where $n=256$. Assuming the same pipelined hardware implementation approach and shot size $Q_{\mathrm{s}}=512$ for ORDEPT and ORBGRAND, \eqref{eq:ClockCycle} is also applicable to the ORDEPT decoder. Moreover, due to the fact that the required $Q_{\mathrm{max}}$ for ORDEPT is 16 times smaller than that of ORBGRAND for $\mathrm{BCH}(256,239)$, ORDEPT is able to decrease the latency caused by the number of shots from 32 (in ORBGRAND) to $2$ clock cycles. In addition $\lceil \log_2( C_{\textrm{max}} )\rceil=2$ extra clock cycles for computing the analog weights of the collected candidate codewords and selecting the smallest weight have been added to the ORDEPT cycle count. 
\begin{table}
    \caption{Comparison of the projected latency in terms of clock cycles of ORDEPT and ORBGRAND implementation \cite{condo2022fixed} for $\mathrm{BCH}(256,239)$.}
    \centering
    \begin{tabular}{ccccc}
    \toprule
        & \multicolumn{1}{c}{ORDEPT} & \multicolumn{1}{c}{ORDEPT} & \multicolumn{1}{c}{ORBGRAND} & \multicolumn{1}{c}{ORBGRAND} \\
        {} & with & with & with & with \\
        {} & $Q_s=512$ & $Q_s=256$ & $Q_s=512$ & $Q_s=256$ \\
        \midrule
        latency [cc] & $14$& $16$& $42$ & $74$  \\
        \bottomrule
    \end{tabular}
    \label{table:2}
    
\end{table}

The estimated clock cycle numbers for ORDEPT and ORBGRAND are given in Table~\ref{table:2}. The table clearly demonstrates the advantage of the proposed ORDEPT in terms of latency. Besides the latency, the required number of shots also translates into the area occupied by the circuit. Finally, we note that ORBGRAND needs a smaller number of clock cycles than Chase II with $p\geq2$ to decode a codeword~\cite{condo2022fixed}. Hence ORDEPT has even larger advantage over Chase~II with $p\geq 2$ in terms of complexity.


\begin{figure}[!t]
\setlength{\unitlength}{1mm}
   \begin{picture}(-10,82)(0,0)
   \put(-3,0){\includegraphics[scale=0.127]{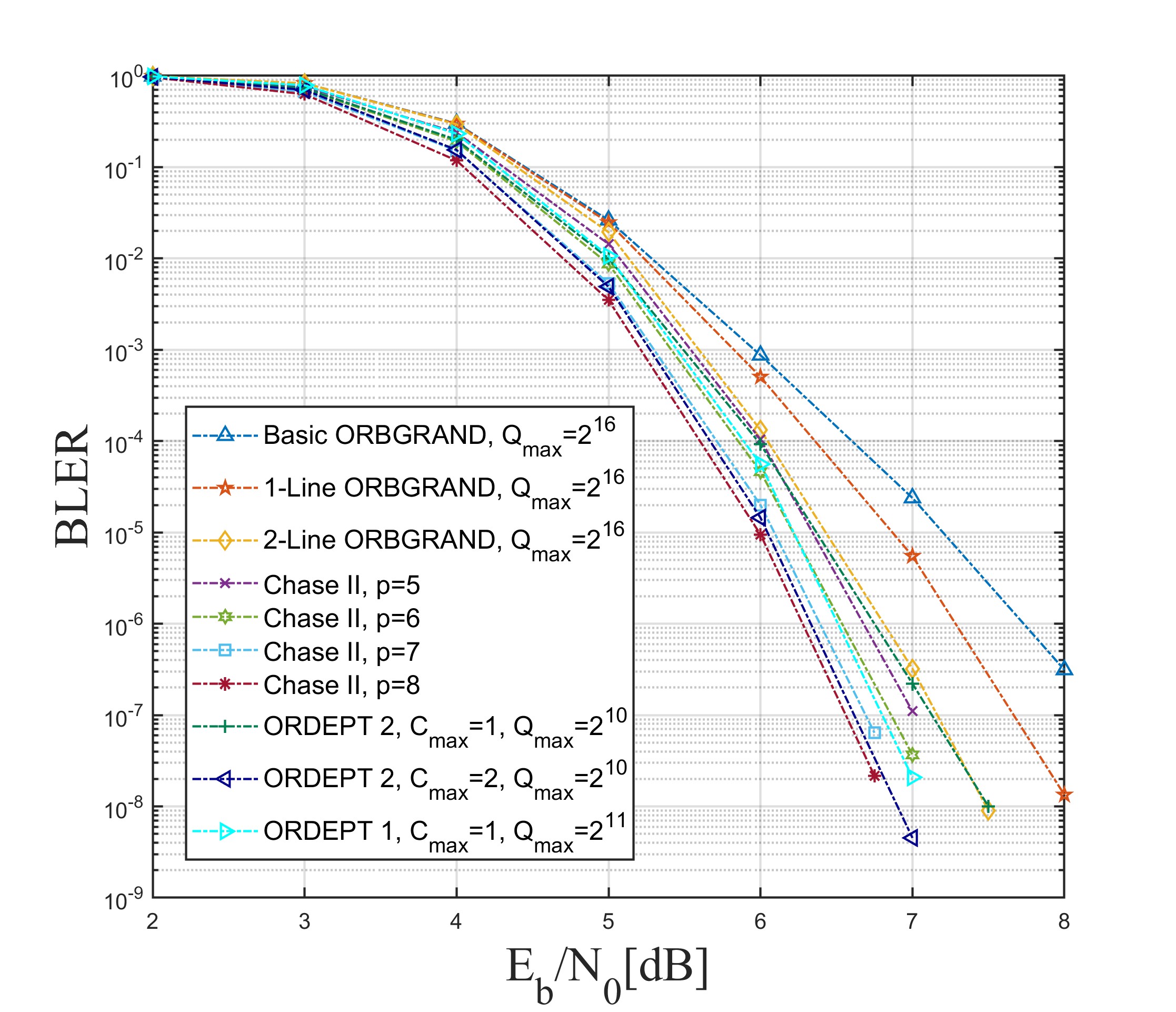}}
\end{picture}
\caption{Performance evaluation of ORDEPT in comparison with existing decoders for $\mathrm{BCH}(256,231)$.}
   \label{Fig:BLERBCH256231}
\end{figure}

In Fig.~\ref{Fig:BLERBCH256231}, we compare BLER performances of ORDEPT1 and ORDEPT2 (i.e. regular ORDEPT and the ORDEPT2 generalization proposed in Section~\ref{subsec:ORDEPTBCH}) with that of Chase II and various ORBGRAND versions for the 3-error-correcting $\mathrm{BCH}(256,231)$ code. Notably, the maximum number of queries $Q_{\mathrm{max}}$ reveals that both ORDEPT2 and ORDEPT exhibit reduced latency, that is $64$ and $32$ times smaller than the latency of ORBGRAND respectively. Furthermore, ORDEPT with $C_\mathrm{max}=1$ slightly outperforms the 2-line version of ORBGRAND. Specifically, at a BLER level of $2\times10^{-8}$, ORDEPT with $C_\mathrm{max}=1$ achieves a gain of approximately $0.5$ dB over  2-line ORBGRAND and a gain of around $1$ dB over 1-line ORBGRAND. For $Q_\mathrm{max}=2^{11}$, on the other hand, ORDEPT is unable to find the second candidate codeword in most cases and, therefore, $C_\mathrm{max}=2$ does not lead to a performance improvement. 

In the case of ORDEPT2, it is evident that with $C_\mathrm{max}=1$, it marginally outperforms the 2-line version of ORBGRAND and achieves a gain of approximately $0.5$ dB over the 1-line ORBGRAND at a BLER level of $10^{-8}$. Unlike ORDEPT, which faces challenges due to the increased distance between the codewords in $\mathrm{BCH}(256,231)$ compared to $\mathrm{BCH}(256,231)$, ORDEPT2 with $C_\mathrm{max}=2$ is capable of identifying a second candidate codeword. This capability results in a gain of more than $0.5$ dB over ORDEPT2 with $C_\mathrm{max}=1$ at a BLER level of $10^{-8}$. The performance of ORDEPT2 with $C_\mathrm{max}=2$ falls between that of Chase II with $p=7$ and Chase II with $p=8$, which find $128$ and $256$ codewords using full $3$ error correction hard-decision decoding respectively.


\begin{figure}[!t]
\setlength{\unitlength}{1mm}
   \begin{picture}(-10,82)(0,0)
   \put(-4,0){\includegraphics[scale=0.125]{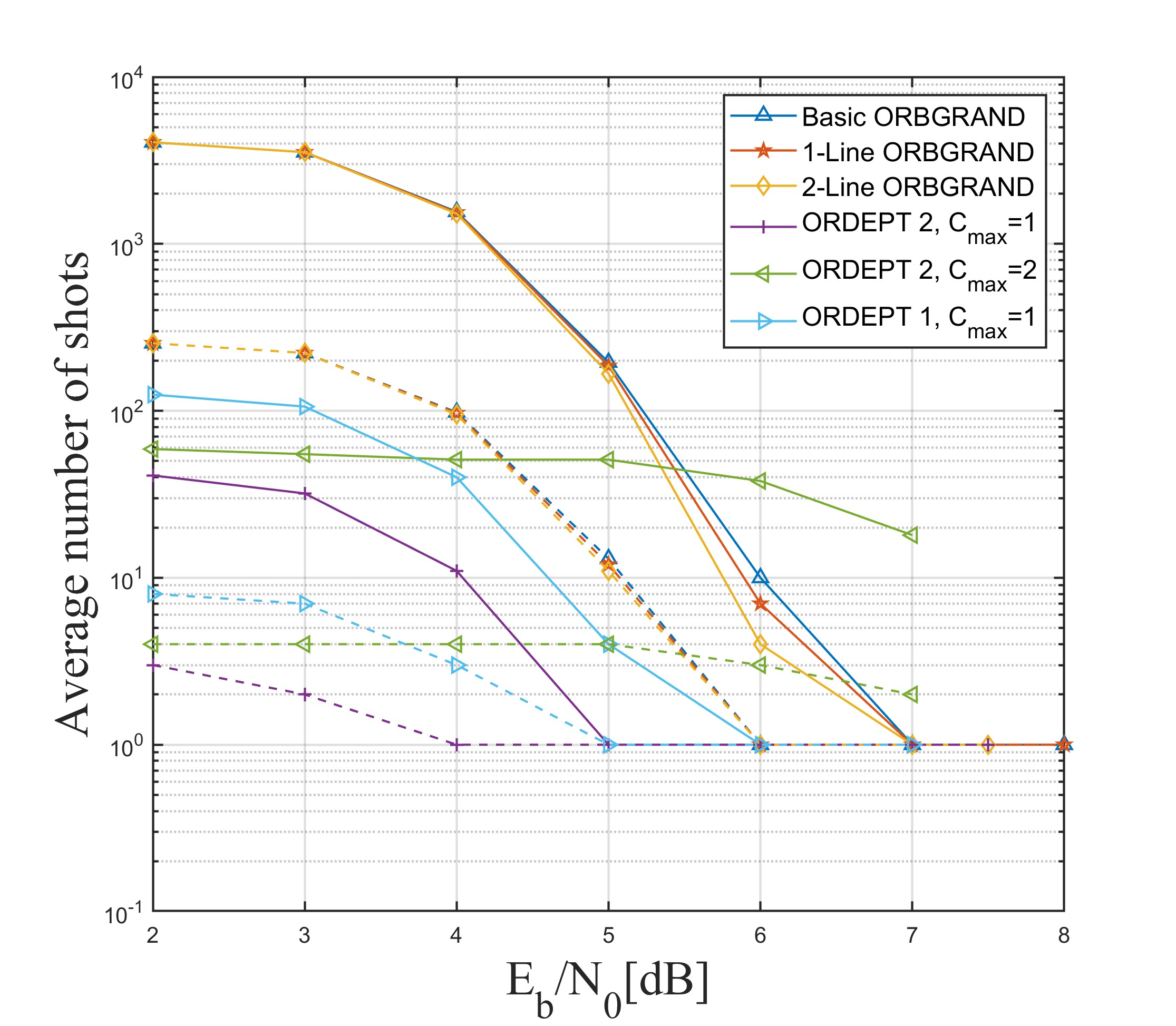}}
\end{picture}
\caption{The average number of shots to decode a codeword with shot size 16 (solid lines) and shot size 256 (dashed lines) using ORDEPT and ORBGRAND variants for $\mathrm{BCH}(256,231)$.}
   \label{Fig:ComplexityBCH256231}
\end{figure}


Fig.~\ref{Fig:ComplexityBCH256231} shows the average number of shots required for ORDEPT and ORBGRAND variants to decode the $\mathrm{BCH}(256,231)$ code. The figure demonstrates that for the shot size $Q_s=256$, the ORBGRAND variants require at least $100$ shots for $\mathrm{SNR}\leq 4$~dB, $10$ shots for $\mathrm{SNR}=5$ dB and $1$ shot for $\mathrm{SNR}\geq 6$~dB. ORDEPT variants however need less than 10 shots for all the SNRs.

\begin{figure}[!t]
\setlength{\unitlength}{1mm}
   \begin{picture}(0,77)(0,0)
   \put(0,0){\includegraphics[scale=0.156]{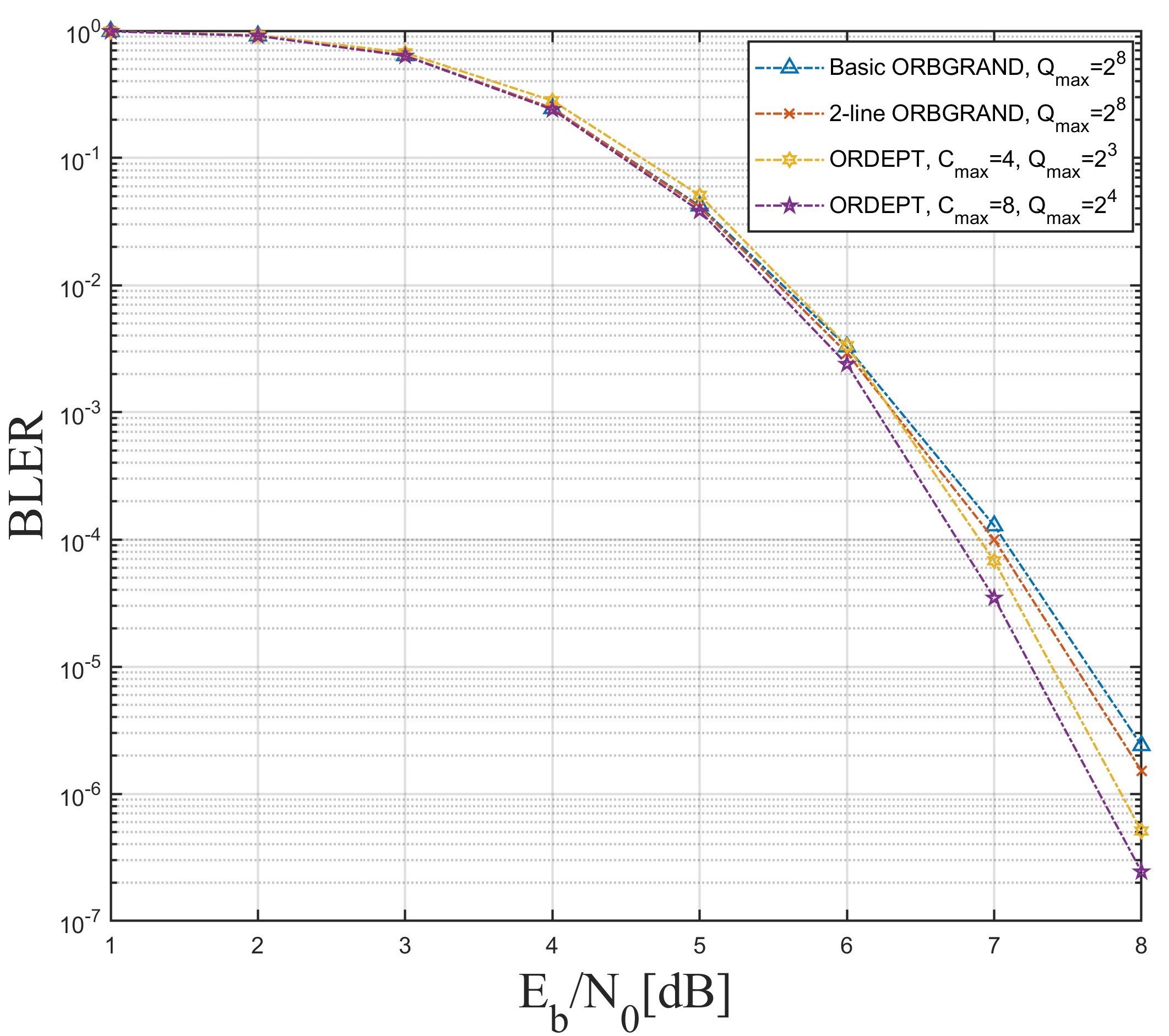}}
\end{picture}
\caption{Performance evaluation of ORDEPT in comparison with ORBGRAND for $\mathrm{CRC}(128,120)$.}
   \label{Fig:BLERCRC128}
\end{figure}
\begin{figure}[!t]
\setlength{\unitlength}{1mm}
   \begin{picture}(0,77)(0,0)
   \put(-5,0){\includegraphics[scale=0.13]{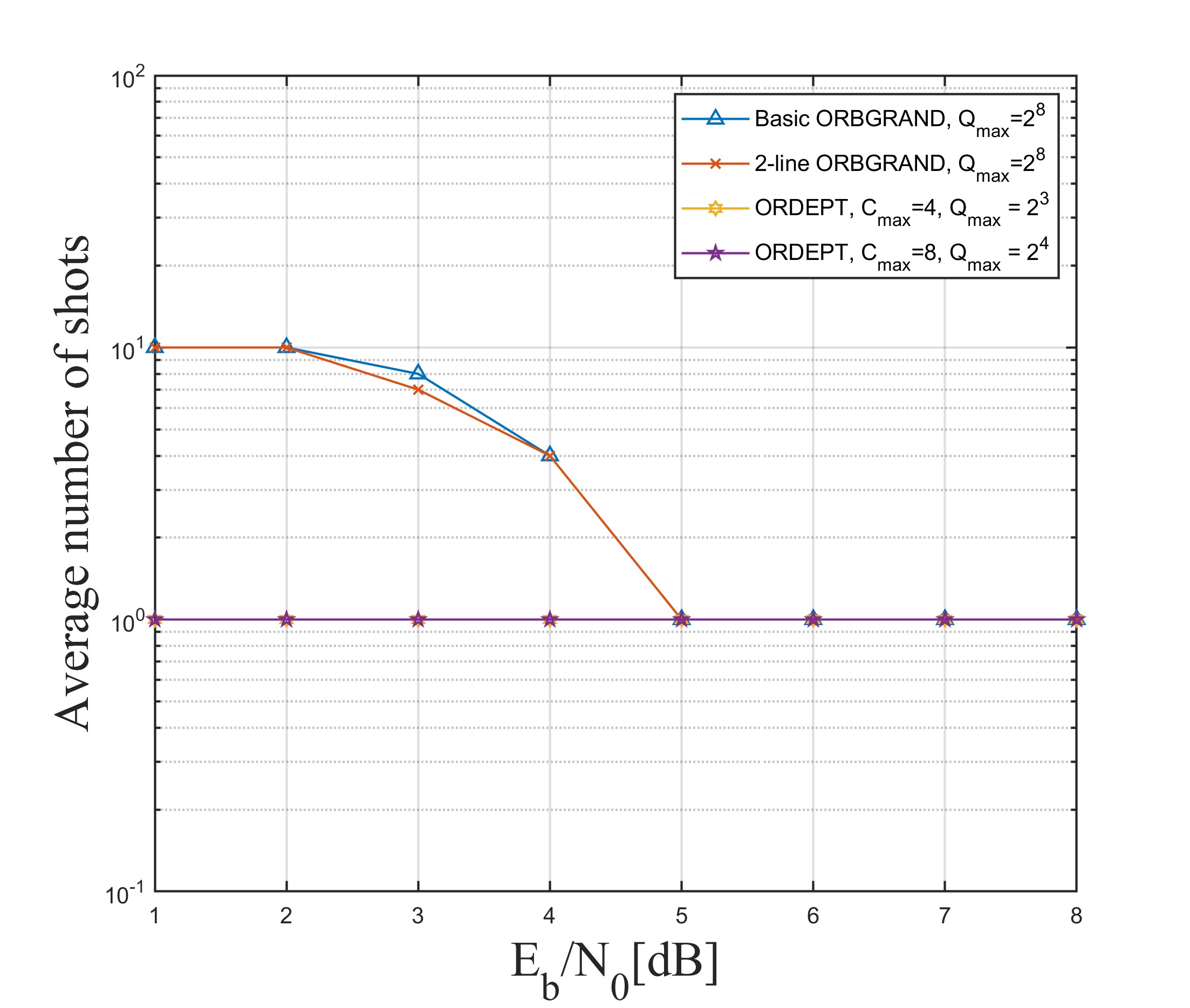}}
\end{picture}
\caption{The average number of shots to decode a codeword with shot size $16$ (solid lines) and shot size $256$ (dashed lines) using ORDEPT and ORBGRAND variants for $\mathrm{CRC}(128,120)$.}
   \label{Fig:ComplexityCRC128}
\end{figure}

Fig.~\ref{Fig:BLERCRC128} presents a comparison of BLER performance between the ORDEPT decoder and ORBGRAND variants applied to $\mathrm{CRC}(128,120)$. The reduced minimum distance of the code allows for reductions in $Q_{\mathrm{max}}$ values for both ORDEPT and ORBGRAND variants to $2^3$ and $2^8$, respectively. As observed from the differences in $Q_{\mathrm{max}}$, the utilization of an analytic function enables the ORDEPT decoder to achieve the latency values that are $16$ to $32$ times smaller compared to the ORBGRAND variants when applied to $\mathrm{CRC}(128,120)$. Furthermore, Fig.~\ref{Fig:BLERCRC128} illustrates that ORDEPT with $\mathrm{C}_{\mathrm{max}}=8$ achieves approximately a $0.2$ dB improvement over the 2-line ORBGRAND variant at BLER level of $10^{-4}$. In Fig.~\ref{Fig:ComplexityCRC128}, the average number of shots required for the decoding of $\mathrm{CRC}(128,120)$ is shown. The results clearly indicate that, regardless of the SNR, ORDEPT consistently demands only a single shot for successful decoding, while ORBGRAND necessitates multiple shots for SNR below $5$~dB.



\begin{figure}[!t]
\setlength{\unitlength}{1mm}
   \begin{picture}(0,78)(0,0)
   \put(0,0){\includegraphics[scale=0.140]{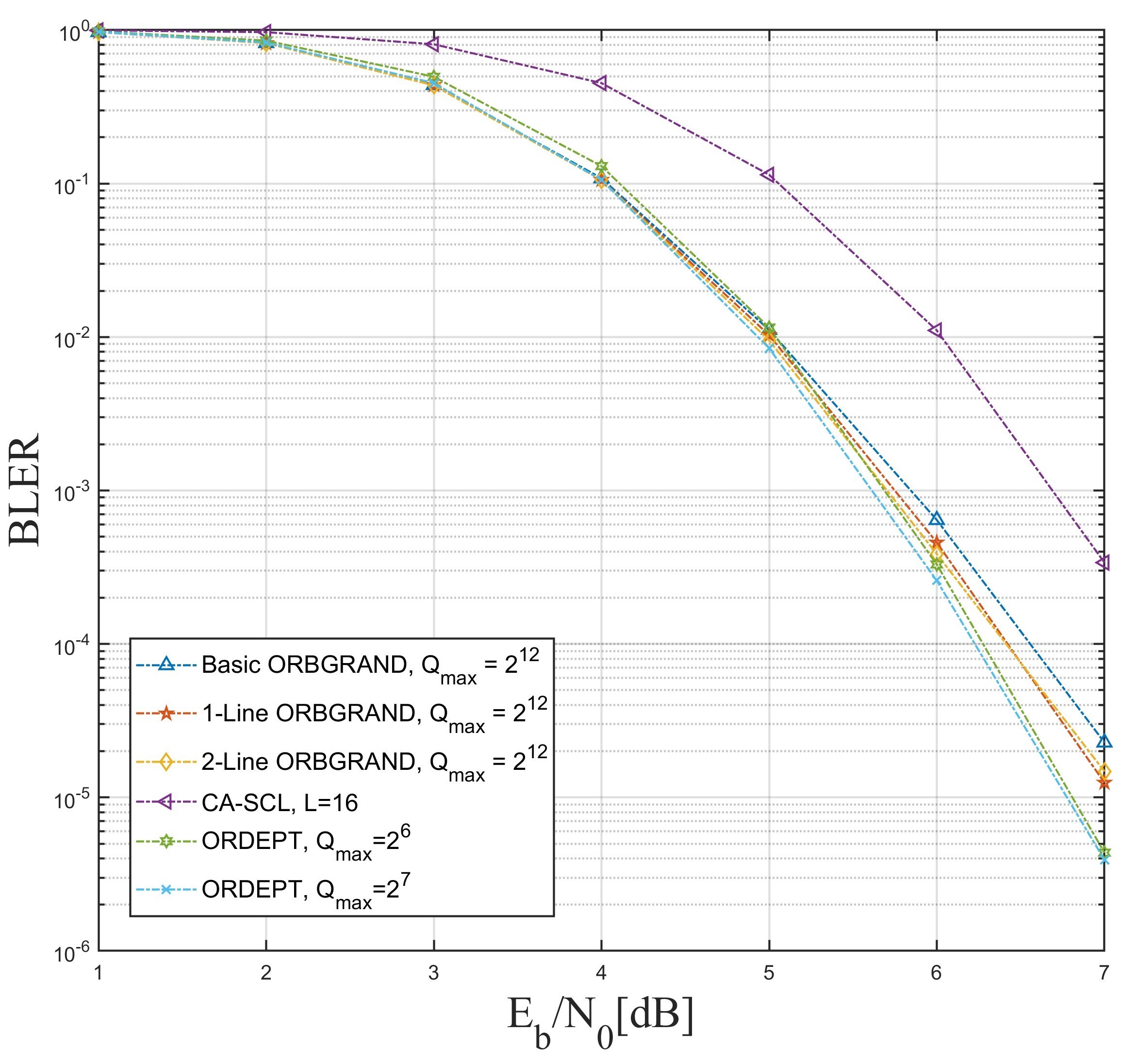}}
\end{picture}
\caption{Performance evaluation of ORDEPT in comparison with ORBGRAND decoders for $\mathrm{CA-Polar}(128,105+11)$.}
   \label{Fig:BLERCAPolar128116}
\end{figure}

\begin{figure}[!t]
\setlength{\unitlength}{1mm}
   \begin{picture}(0,77)(0,0)
   \put(-5,0){\includegraphics[scale=0.13]{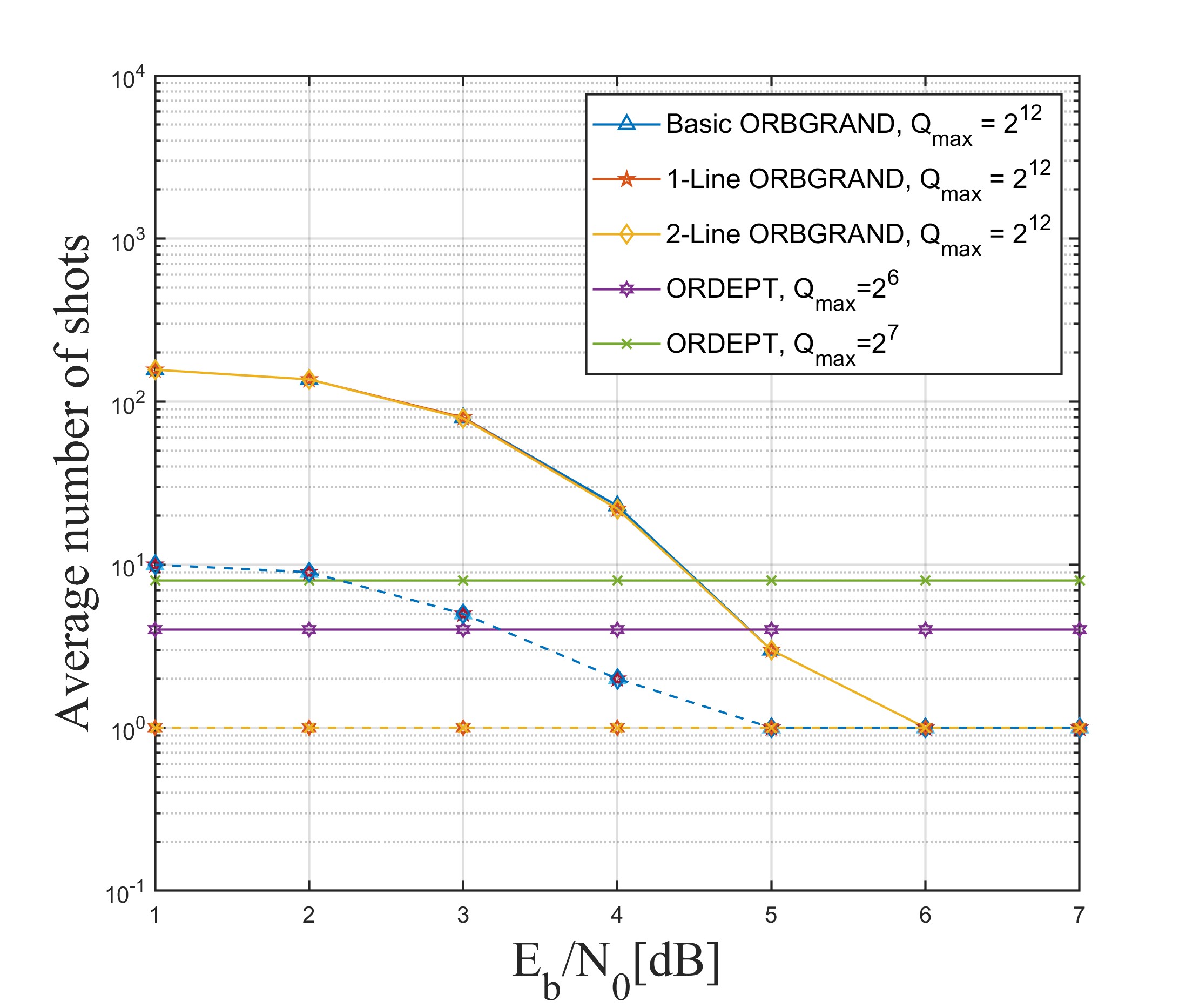}}
\end{picture}
\caption{The average number of shots to decode a codeword with shot size $16$ (solid lines) and shot size $256$ (dashed lines) using ORDEPT and ORBGRAND variants for $\mathrm{CA-Polar}(128,105+11)$.}
   \label{Fig:ComplexityCAPolar128116}
\end{figure}

In Fig.~\ref{Fig:BLERCAPolar128116}, we conduct a performance comparison of ORDEPT with various ORBGRAND variants and the CA-SCL decoder applied to the $\mathrm{CA-Polar}(128,116)$ code. This code, designed for 5G NR up-link control channels, contains $22$ parity bits in addition to $11$ CRC bits. The results in the figure indicate that ORDEPT outperforms the CA-SCL decoder by a margin of $1$~dB and surpasses ORBGRAND by $0.1$~dB at a BLER level of $3\times10^{-4}$. Furthermore, it is worth noting that the ORDEPT decoder exhibits a latency that is $32$ times smaller compared to the ORBGRAND variants. Fig.~\ref{Fig:ComplexityCAPolar128116} indicates that ORDEPT needs less shots than ORBGRAND variants for lower SNRs but more for higher SNRs above $5$~dB for $Q_\mathrm{s}=16$. However, for a more practical  $Q_\mathrm{s}=256$ ORDEPT always requires the same or smaller amount of shots in comparison to ORBGRAND.


\begin{figure}
    \setlength{\unitlength}{1mm}
    \begin{picture}(0,100)(0,0)
    \put(0,0){\includegraphics[scale=0.10]{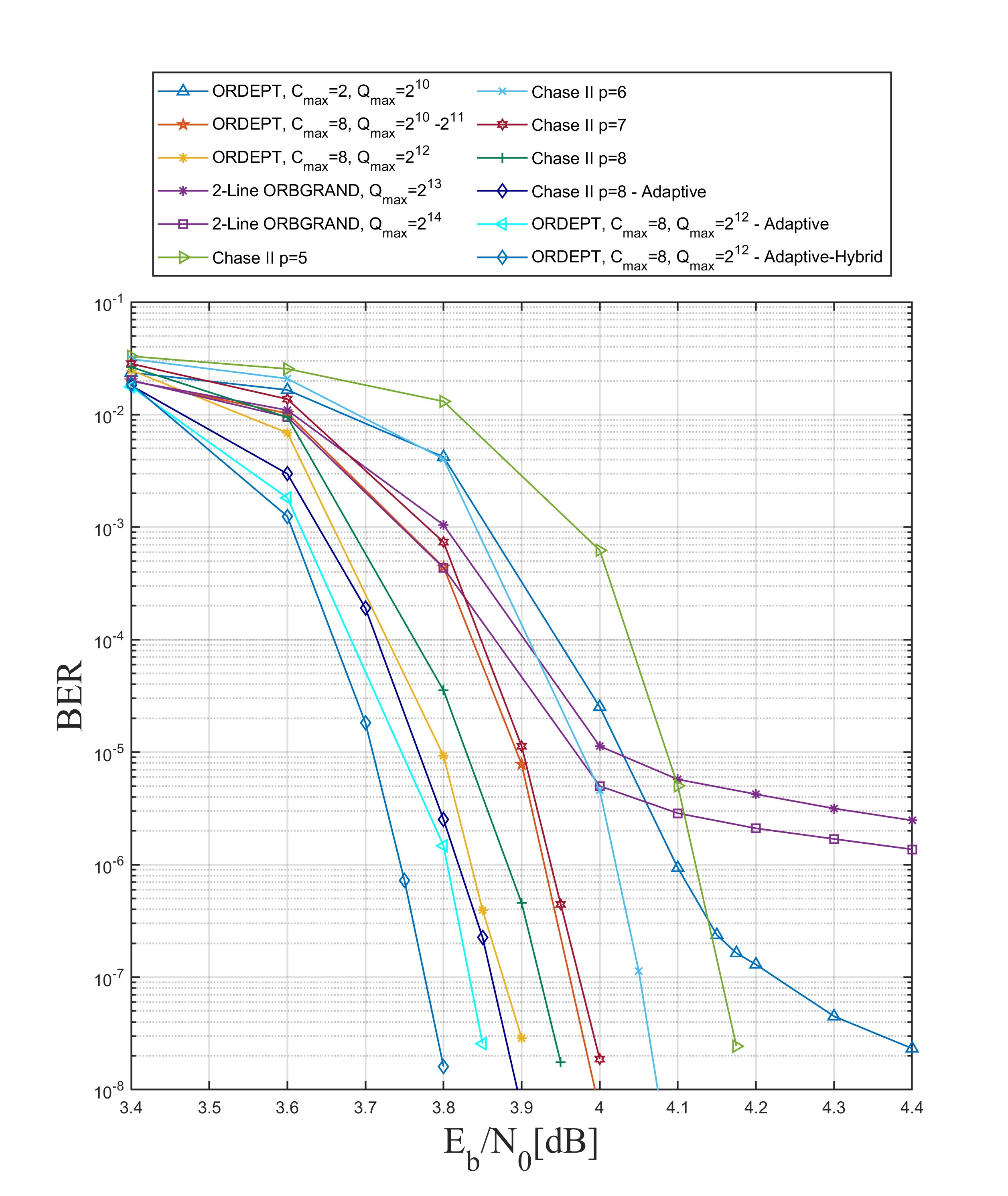}}
    \end{picture}
\caption{BER of a $\mathrm{BCH}(256,239)^2$ product code with soft iterative decoding. ORDEPT, ORBGRAND and Chase II algorithms are used as component decoders.}
\label{Fig:Product_BER}
\end{figure}

The BER performance of a $\mathrm{BCH}(256,239)^2$ product code decoded using various component code decoders is presented in Fig.~\ref{Fig:Product_BER}. Notably, the 2-line version of ORBGRAND (as well as a simple and 1-line variants not presented in the figure) exhibits an error floor, primarily attributed to its limited capability in finding a competitor codeword in over $95\%$ of decoding attempts, even with a high number of queries set at $Q_{\textrm{max}}=2^{14}$. In contrast, ORDEPT consistently provides more than one candidate codeword in over $99\%$ of cases.

ORDEPT with $C_{\mathrm{max}}=2$ exhibits performance which is superior than the performance of Chase II with $p=5$ for BER above $10^{-6}$. However, for lower BER values, the performance of ORDEPT degrades, as having only two candidate codewords becomes insufficient for accurately representing the soft information. In contrast, ORDEPT with $Q_{\mathrm{max}}=2^{11}$ and $C_{\mathrm{max}}=8$ outperforms even the more complex Chase~II $p=7$ decoder which needs to identify $128$ codewords through hard-decision decoding. The figure also depicts the BER performance of Chase II with $p=8$ alongside ORDEPT employing an adaptive method proposed in \cite{deng2023adaptive}. Notably, our proposed adaptive hybrid method for soft information update shows a gain of $0.09$ dB over the adaptive Chase II with $p=8$ at a BER level of $10^{-7}$. It is essential to highlight that this gain from the proposed adaptive hybrid method is attainable solely through ORDEPT, which provides multiple codewords ranked by likelihood.



\begin{figure}
    \setlength{\unitlength}{1mm}
    \begin{picture}(0,82)(0,0)
    \put(-6,0){\includegraphics[scale=0.103]{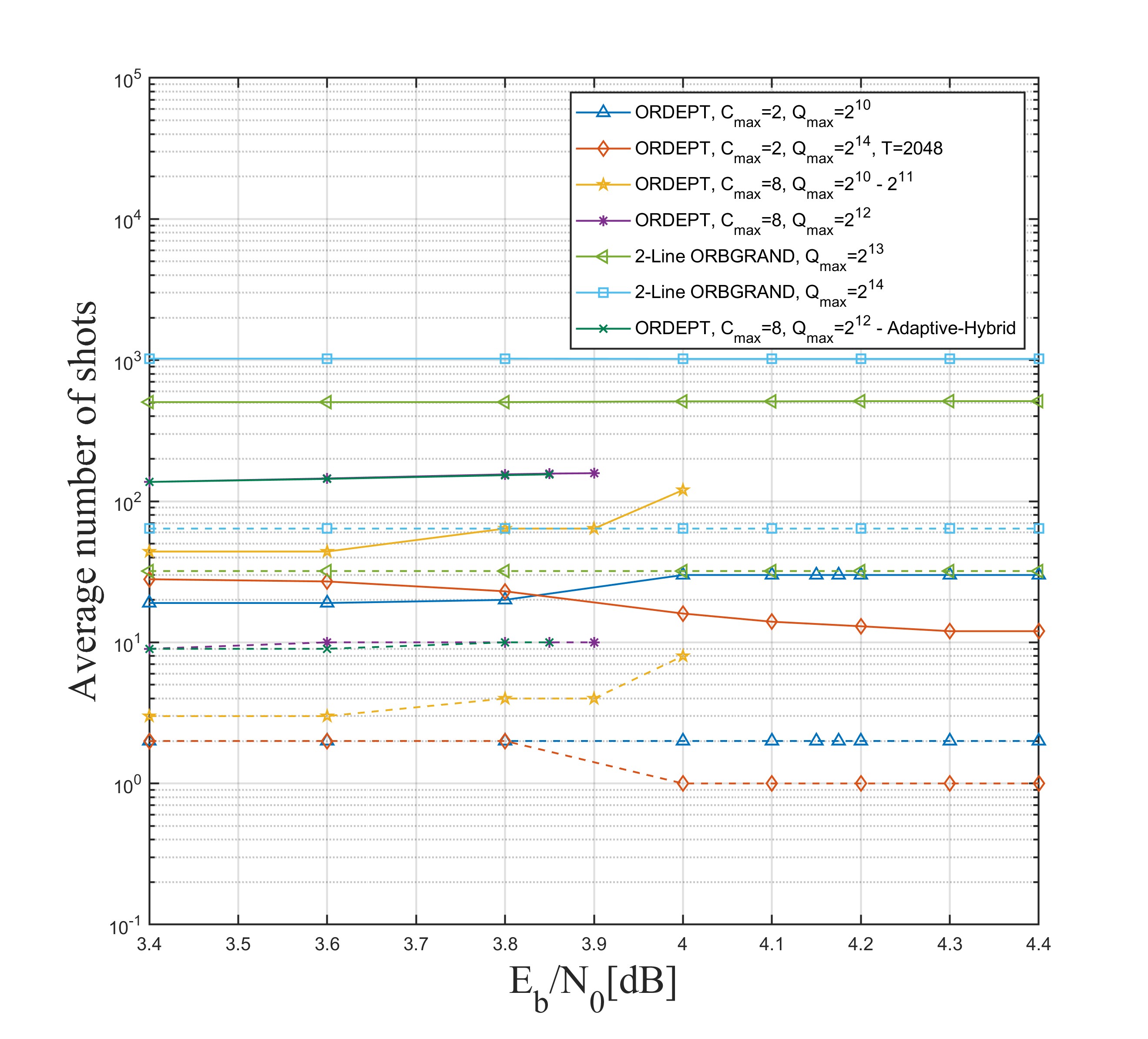}}
    \end{picture}
\caption{The average number of shots per component code for product code decoding using shot size $16$ (solid lines) and shot size $256$ (dashed lines) using ORDEPT and ORBGRAND variants.}
\label{Fig:Product_Complexity}
\end{figure}

Fig.~\ref{Fig:Product_Complexity} presents the average number of shots for the case of product code decoding. The results showcased in this figure highlight the significant complexity reduction achieved by the proposed decoder, surpassing ORBGRAND by an order of magnitude. As it is evident, for shot size $Q_s=512$, while 2-line ORBGRAND with $Q_{\mathrm{max}}=2^{14}$ requires, in average, around $70$ shots per component code, ORDEPT variants need at most 10. The reason for the increase in the last point of yellow curve is that $Q_\mathrm{max}~=~2^{11}$ is used instead of $Q_\mathrm{max}=2^{10}$ since it prevents the occurrence of an error floor. This complexity reduction, coupled with substantial improvements in BER performance, underscores the effectiveness of the proposed decoder.



\begin{figure}[!t]
    \setlength{\unitlength}{1mm}
    \begin{picture}(0,82)(0,0)
    \put(-2,0){\includegraphics[scale=0.485]{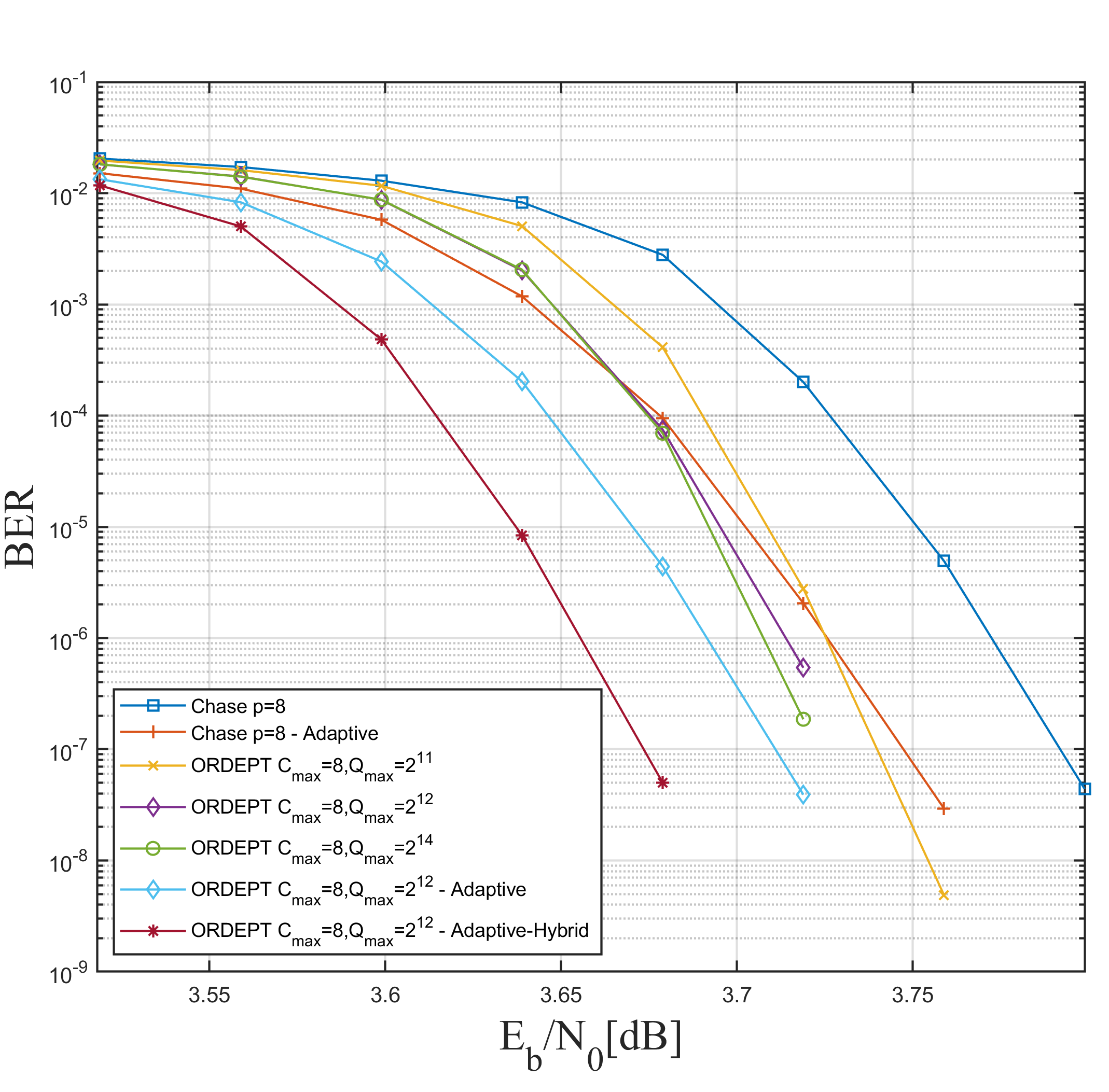}}
    \end{picture}
\caption{BER for iterative decoding of the oFEC code (3 soft and 2 hard iterations). ORDEPT and Chase II type algorithm are used as component decoders.}
\label{fig:oFEC_BER}
\end{figure}

BER performances of the $\mathrm{BCH}(256,239)$ based oFEC codes with various decoders are shown in Fig.~\ref{fig:oFEC_BER}. Similarly to the product code case, adaptive ORDEPT with $C_{\textrm{max}}=8$  outperforms the complex Chase II $p=8$ decoder on the oFEC codes. Moreover, the gain of the proposed adaptive-hybrid method over the Chase II $p=8$ is even more significant and is around $0.075$ dB at BER level $10^{-6}$. 
While the parameter $C_\mathrm{max}=8$ remains unchanged, the error correction performance slowly improves as $Q_\mathrm{max}$ increases. According to the BER plot, the performances of the $C_\mathrm{max}=8$, $Q_\mathrm{max}=2^{12}$ ORDEPT and the $C_\mathrm{max}=8$, $Q_\mathrm{max}=2^{14}$ ORDEPT are quite close to each other. The reason for this phenomenon is that the specified number of candidate codewords $C_\mathrm{max}$ is reached faster than the total number of queries $Q_\mathrm{max}$ and saturates the error correction performance. The average number of shots, for the shot size $Q_\mathrm{s}=256$, required by ORDEPT for all presented versions is approximately $10$ across all SNRs.



\section{Conclusion} \label{sec:conclusion}
A new decoding algorithm for binary block codes called ORDEPT that is based on completion of pre-defined partial error patterns ranked in the order of likelihood is proposed. 
A generalization of the algorithm that completes partial error patterns with more than one error position is also presented. The numerical results demonstrate that ORDEPT delivers performance improvements as well as complexity and latency reduction, compared to the state-of-the-art algorithms.  For the case of product-like codes, ORDEPT, unlike ORBGRAND variants,  is capable of finding multiple candidate codewords and effectively competing with conventional Chase II decoders that necessitate multiple full hard-decision decoding attempts.



\balance
\printbibliography

\end{document}